\documentclass[aps,pra,twocolumn, showpacs]{revtex4}

\usepackage{amsmath}
\usepackage{amsfonts,amssymb,latexsym,amsthm}
\usepackage{amscd}
\usepackage{feynmp}
\usepackage{fancybox}
\usepackage{bm}
\usepackage{times}

\usepackage{dsfont}
\usepackage{textcomp}

\usepackage[english]{babel}  

\usepackage{ifvtex}
\usepackage{ifpdf}
\ifvtexpdf\pdftrue\fi
\ifpdf
\usepackage[pdftex]{graphicx}            
\ifvtex\relax
\else
\DeclareGraphicsExtensions{.pcx}
\fi
\else
\usepackage[dvips]{graphicx}
\DeclareGraphicsExtensions{.eps.gz,.eps}
\fi
\begin{document}


\newcommand{\ie}{{i.{\thinspace}e.\,}}
\newcommand{\bg}{\begin{gather}}
\newcommand{\eg}{\end{gather}}

\newcommand{\ba}{\begin{align}}
\newcommand{\ea}{\end{align}}

\newcommand{\be}{\begin{equation}}
\newcommand{\ee}{\end{equation}}
\newcommand{\bea}{\begin{eqnarray}}
\newcommand{\eea}{\end{eqnarray}}
\newcommand{\ket}[1]{\left| #1\right \rangle}
\newcommand{\bra}[1]{\left \langle #1\right|}
\newcommand{\twoket}[2]{\left| #1\right \rangle
                        \otimes
                        \left| #2\right \rangle }
\newcommand{\twobra}[2]{\left \langle #1\right|
                        \otimes
                        \left \langle #2\right|}
\newcommand{\scal}[2]{\langle #1| #2 \rangle}
\newcommand{\av}[1]{\langle #1 \rangle}
\newcommand{\aval}[1]{\langle #1 
  \rangle_{\{\al{},\ald{},\fs{},\ms{}\}}}
\newcommand{\avnull}[1]{\langle #1
        \rangle_{\{0,0,\fs{},\ms{}\}}}
\newcommand{\trace}[1]{\textrm{Tr}\left\{ #1 \right\}}
\newcommand{\comut}[2]{[#1,\,#2\,]}

\newcommand{\al}[1]{\alpha_{#1}^{\!\!\phantom{\ast}}}
\newcommand{\ald}[1]{\alpha_{#1}^\ast}
\newcommand{\alp}[1]{ \alpha_{{#1}^\prime}^{\!\!\phantom{\ast}}}
\newcommand{\aldp}[1]{\alpha_{{#1}^\prime}^\ast}
\newcommand{\alpp}[1]{ \alpha_{{#1}^\ppr}^{\!\!\phantom{\ast}}}
\newcommand{\aldpp}[1]{\alpha_{{#1}^\ppr}^\ast}

\newcommand{\piGP}[1]{\Pi_{{\cal{#1}}}}

\newcommand{\Gg}{G^{>}}
\newcommand{\Gk}{G^{<}}

\newcommand{\Upg}[1]{\Upsilon_{\cal{#1}}^{>}}
\newcommand{\Upk}[1]{\Upsilon_{\cal{#1}}^{<}}

\newcommand{\Gag}[1]{\Gamma_{\cal{#1}}^{>}}
\newcommand{\Gak}[1]{\Gamma_{\cal{#1}}^{<}}

\newcommand{\sigN}{\Sigma_{\cal{N}}}
\newcommand{\sigA}{\Sigma_{\cal{A}}}

\newcommand{\mf}{\ket{\al{}}}
\newcommand{\mfd}{\bra{\al{}}}

\newcommand{\paul}[1]{\sigma_{{#1}}}

\newcommand{\aop}[1]{{ \hat{a}_{#1}^{\!\!\phantom{\dag}}}}
\newcommand{\aopd}[1]{{\hat{a}_{#1}^{\dag}}}
\newcommand{\ephn}{e^{i\hn \tau}}
\newcommand{\emhn}{e^{-i\hn \tau}}

\newcommand{\Vop}{\hat{V}}
\newcommand{\gamopal }{\hat{\gamma}_{\alpha}}
\newcommand{\gamopi }{\hat{\gamma}_{i}}
\newcommand{\gamopbet}{\hat{\gamma}_{\beta }}
\newcommand{\gamopj}{\hat{\gamma}_{j}}

\newcommand{\gamopset}[1]{\{  \hat{\gamma}_{#1}|\,#1 \in {\cal I}\}}

\newcommand{\gamal }{\gamma_{\alpha}}
\newcommand{\gami }{\gamma_{i}}
\newcommand{\gambet}{\gamma_{\beta }}
\newcommand{\gamj}{\gamma_{j}}

\newcommand{\Upsq}[1]{{\Upsilon}^{#1}_{ \{\fs{},\ms{},\ns{}\}}}
\newcommand{\Lamsq}[1]{{\Lambda}^{#1}_{\{\fs{},\ms{},\ns{}\}}}

\newcommand{\Gcoll}[1]{{\Gamma_{#1}}}

\newcommand{\bgamal }{\overline{\gamma}_{\alpha}}
\newcommand{\bgami }{\overline{\gamma}_{i}}
\newcommand{\bgambet}{\overline{\gamma}_{\beta }}
\newcommand{\bgamj}{\overline{\gamma}_{j}}

\newcommand{\sigam}[1]{{\mathbf{\sigma}}_{\{\gamma #1\}}}

\newcommand{\brho}{{\mathbf{\rho}}}

\newcommand{\signal}{{\mathbf{\sigma}}_{
        \{\al{},\ald{},\fs{},\ms{},\ns{}\}}^{(0)}}
\newcommand{\sign}[1]{{\mathbf{\sigma}}_{\{\gamma #1\}}^{(0)}}

\newcommand{\signSchroedinger}[2]{
        \sigma_{ \{\overline{\gamma}(#1;\{\gamma #2\})\}}^{(0)}}

\newcommand{\sigo}[1]{{\mathbf{\sigma}}_{\{\gamma #1\}}^{(1)}}
\newcommand{\sigoSchroedinger}[2]{
        \sigma_{ \{\overline{\gamma}(#1;\{\gamma #2\})\}}^{(1)}}

\newcommand{\brhoref}{{\mathbf{\rho}}_{\{ \alpha, \alpha^\ast, f \}}^{(0)}}
\newcommand{\Undag}[1]{\left. {{\widehat{U}}_{\{\gamma\}}^{(0)}}\!\!\right. 
^{\dagger}\!\!(#1)}
\newcommand{\Un}[1]{\widehat{U}_{\{\gamma\}}^{(0)}(#1)}

\newcommand{\bx}{{\mathbf{x}}}
\newcommand{\bp}{{\mathbf{p}}}
\newcommand{\by}{{\mathbf{y}}}
\newcommand{\bz}{{\mathbf{z}}}
\newcommand{\bal}{{\mathbf{\alpha}}}
\newcommand{\hng}[1]{\widehat{H}^{(0)}_{\{\gamma #1\}}}
\newcommand{\hngSchroedinger}[2]{
        \widehat{H}_{ \{\overline{\gamma}(#1;\{\gamma #2\})\}}^{(0)}}
\newcommand{\hog}[1]{\widehat{H}^{(1)}_{\{\gamma #1\}}}
\newcommand{\hogSchroedinger}[2]{
        \widehat{H}_{ \{\overline{\gamma}(#1;\{\gamma #2\})\}}^{(1)}}
\newcommand{\Qog}[1]{\widehat{Q}^{(1)}_{\{\gamma #1\}}}
\newcommand{\QogSchroedinger}[2]{
        \widehat{Q}_{ \{\overline{\gamma}(#1;\{\gamma #2\})\}}^{(1)}}
\newcommand{\hn}{\widehat{H}^{(0)}}
\newcommand{\ho}{\widehat{H}^{(1)}}
\newcommand{\astruc}[1]{{{\mathcal{A}}_{\{{#1}\}}}}
\newcommand{\psiop}[1]{\hat{a}_{#1}^{\phantom \dag}}
\newcommand{\psiopd}[1]{\hat{a}_{#1}^\dag}
\newcommand{\kpsiop}[1]{\hat{\psi}_{#1}}

\newcommand{\psum}{\sum_{1^\prime 2^\prime 3^\prime 4^\prime}}
\newcommand{\ppsum}{\sum_{1^\ppr 2^\ppr 3^\ppr 4^\ppr}}
\newcommand{\ppr}{{\prime \prime}
}
\newcommand{\fs}[1]{\tilde{f}_{#1}}

\newcommand{\fsop}[1]{{\,\,\hat{\!\!\tilde{f}}_{#1}}}
\newcommand{\fsT}[1]{\tilde{f}_{#1}^{\,\top}}
\newcommand{\fspo}[1]{(\mathds{1}+\tilde{f})_{#1}}
\newcommand{\fspoT}[1]{(\mathds{1}+\tilde{f})_{#1}^{\ast}}
\newcommand{\ms}[1]{\widetilde{m}_{#1}}

\newcommand{\msop}[1]{{\hat{\widetilde{m}}}_{#1}}
\newcommand{\ns}[1]{\widetilde{n}_{#1}}

\newcommand{\nsop}[1]{\hat{\widetilde{n}}_{#1}}
\newcommand{\fse}[1]{\tilde{f}_{\varepsilon_{#1}}}
\newcommand{\fsep}[1]{\tilde{f}_{\varepsilon_{{#1}^\prime}}}
\newcommand{\fsepp}[1]{\tilde{f}_{\varepsilon_{{#1}^\ppr}}}

\newcommand{\fc}[1]{f^{(c)}_{#1}}
\newcommand{\mc}[1]{m^{(c)}_{#1}}
\newcommand{\nc}[1]{n^{(c)}_{#1}}
\newcommand{\ft}[1]{f_{#1}}
\newcommand{\mt}[1]{m_{#1}}
\newcommand{\nt}[1]{n_{#1}}

\newcommand{\phiunpr}{\phi^{1 2 3 4}}
\newcommand{\phipr}{\phi^{1\,2^\prime 3^\prime 4^\prime}}
\newcommand{\phippr}[1]{\phi^{1^\ppr 2^\ppr 3^\ppr 4^\ppr}_{#1}}

\newcommand{\phiprb}{\phi^{\bar{1}\,\bar{2}^\prime \bar{3}^\prime 
\bar{4}^\prime}}
\newcommand{\phipprb}[1]{\phi^{\bar{1}^\ppr \bar{2}^\ppr \bar{3}^\ppr 
\bar{4}^\ppr}_{#1}}

\newcommand{\phireduc}[4]{\phi^{\bar{#1} \bar{#2} \bar{#3} \bar{#4}}}
\newcommand{\Yangul}[4]{Y^{{\underline{#1}}\, {\underline{#2}}\, 
{\underline{#3}}\, {\underline{#4}}}}
\newcommand{\gangul}[4]{g^{{\bar{#1}}\, {\bar{#2}}\, {\bar{#3}}\, {\bar{#4}}}}
\newcommand{\gangulprb}{g^{\bar{1} \bar{2}^\prime \bar{3}^\prime
\bar{4}^\prime}}

\newcommand{\eps}[1]{\varepsilon_{{#1}}}
\newcommand{\epsp}[1]{\varepsilon_{{#1}^\prime}}
\newcommand{\epspp}[1]{\varepsilon_{{#1}^{\prime \prime}}}

\newcommand{\dd}[1]{{\partial_{ #1}}} 

\newcommand{\Uc}{\mathcal{U}_{\fc{}}}
\newcommand{\Usq}{\mathcal{U}_{\fs{}}}
\newcommand{\Vc}{\mathcal{V}_{\mc{}}}
\newcommand{\Vsq}{\mathcal{V}_{{\ms{}}}}
\newcommand{\Vm}{\mathcal{V}_{(\mc{}+\ms{})}}
\newcommand{\Ltwo}[1]{L_{\{\al{},\ald{},\fs{},\ms{},\ns{}\}}^{(2)}[#1]}

\newcommand{\BE}{Bose-{E}instein }
\newcommand{\FD}{Fermi-Dirac }

\title{Ground state correlations in a trapped quasi one-dimensional Bose Gas}
\author{R. Walser} \email{Reinhold.Walser@physik.uni-ulm.de}
\affiliation{Abteilung f\"ur Quantenphysik, Universit\"at Ulm, D-89069, Ulm,
  Germany}

\date{\today}

\begin{abstract}
  We review the basic concepts of a non-equilibrium kinetic theory of a
  trapped bosonic gas. By extending the successful mean-field concept of the
  Gross-Pitaevskii equation with the effects of non-local, two particle
  quantum correlations, one obtains a renormalized binary interaction and
  allows for the dynamic establishment of non-classical many-particle quantum
  correlations. These concepts are illustrated by self-consistent numerical
  calculations of the first and second order ground state quantum correlations
  of a harmonically trapped, quasi one-dimensional bosonic gas. We do find a
  strong suppression of the density fluctuations or, in other words, an
  enhanced number squeezing with decreasing particle density.
\end{abstract}


\maketitle
\section{Introduction}
Already the state of a single quantum mechanical particle can exhibit an
amazingly complexity. Without the restriction to one space dimension or by
exploiting additional symmetries, it is virtually impossible even to visualize
the information that is encoded in a single particle wave function.  Even more
so, this situation gets quickly out-off hand by adding more interacting
particles to the system.  While the new field of quantum information science
is exactly trying to accomplish this task of describing and preparing
many-particle entanglement, it remains a formidable challenge. In contrast,
many-particle physics is blessed with serendipity as the astronomic growth in
the dimensionality of the state space leads to a great reduction of
complexity. No macroscopic physical phenomenon does rely on a miniscule detail
of a particular quantum state as long as the whole ensemble exhibits a certain
macroscopic characteristic.

The past decade of research in degenerate atomic gases has produced an amazing
wealth of condensed many-particle phenomena
\cite{stringarireview,leggett401,southwell,pethick02}: Bose-Einstein
condensation \cite{cornellsci,ketterle1195,hulet895}, the creation of vortices
\cite{holland999,matthews99}, Abrikosov lattices \cite{ketterle0401}, the Mott
phase transition \cite{jaksch98,bloch02,Greiner2002b}, the creation
of molecular condensates \cite{regal03,zwierlein04}, the BCS-BEC crossover 
\cite{regal04,chin04} and the one-dimensional 
Tonks-Girardeau gas \cite{bloch04,das02}, most of which have been seen or, at least,
were predicted in other, traditional fields of low temperature physics.
However, observing the original BEC phase transition of a dilute atomic gas in
situ and real space is still a delightful lesson in fundamental physics, as
one can be an eye witness of the establishment of off-diagonal long range
order (ODLRO) \cite{onsager56,yang62} and get an impression of role that is
carried by quantum fluctuations .

The art of performing successful many-particle calculations consists of
picking the right approximation scheme that matches the experimental system on
one hand and that is theoretically tractable on the other hand.  In the
context of weakly correlated dilute Bose gases, the Gross-Pitaevskii (GP)
mean-field picture has been a tremendously rewarding concept and the extension
to incorporate quantum fluctuations dynamically is in principle straight
forward, although it involves in detail some intricate calculations
\cite{beliaev58b,Hohenberg1965,
  zaremba899,rusch599,Stoof1999a,walser599,wachter501,giorgini00}. The
description of the temporal
relaxation \cite{jackson02,bhongale}, the buildup of spatial correlation
functions \cite{Naraschewski996,esslinger02} and the squeezing of 
atomic number density fluctuations \cite{Orzel01} follow, consequently.

However, it has been recognized very early on that the spatial dimensionality
of a system is of utmost importance to its physical behavior. In particular,
it has been proven that the reduction of the available phase space volume
leads to enhanced fluctuations and the absence of ODLRO in one and two
dimensions \cite{mermin66,hohenberg67}. During the last years, this
fascinating observation has received much attention
\cite{shlyapnikov04,andersen02,alkhawaja02,ertmer03} as trapped, inhomogeneous
systems violate the translational invariance as required in
\cite{mermin66,hohenberg67}. This different physical response should also be
reflected by tuning the trap geometry dynamically from
3d$\rightarrow$2d$\rightarrow$ quasi 1d $\rightarrow$ 1d
\cite{Olshanii98,menotti02,das02,bloch04,esslinger04}. The extended mean-field
theory that will be presented in the following, is well suited to describe the
cross-over physics with the exception of the strongly interacting
Tonks-Girardeau regime.

We have arranged this article according to the following outline: In
Sec.~\ref{sectwo}, we will give a basic review of the premises and concepts of
non-equilibrium kinetic theory. In particular, we discuss the physical meaning
of the relevant master variables that are used. We discuss the derivation and
approximations that lead to the the basic self-consistent time-dependent
Hartree-Fock-Bogoliubov (HFB) equations of motion in the absence of collision.
A short review on the mathematical properties of "Bogoliubov-like" symplectic
self energy operators is given, taking special care of the presence or absence
of a zero energy mode.  Sec.~\ref{secres} is devoted to an application of the
general formulation to a quasi one-dimensional trapped bosonic gas.  To
calculate specific numbers, we are assuming in here the typical data of a
${}^{87}$Rb experiment. In particular, we calculate for the zero-temperature
ground state of a gas: the mean-amplitude, the quantum depletion, the pairing
field, as well as the first and second order correlation functions for a full
range of particle numbers $N=(10^0,\ldots,10^5)$. In Sec.~\ref{concout}, we do
draw conclusion and give an outlook to the work in progress. Finally, five
short appendices compile some technical methods or basic statements that were
used in the article.

\section{Collisionless Kinetic equations} 
\label{sectwo}
\subsection{Quantum dynamics}
The kinetic evolution of a trapped atomic gas is described very well by a
dilute gas Hamiltonian \cite{stringarireview}. In the limit of strongly
rarefied atomic gases, it consists primarily of the single particle energy of
atoms in a harmonic trap potential $V_{\text{ho}}$ and the mutual interaction
energy amongst all pairs of atoms which is mediated through a short-range
inter-atomic potential $V_{\text{bin}}$.  Strong collisions between atomic
triples are very unlikely events in this dilute gas limit and can be
disregarded, consequently.  Thus, one finds for the Hamilton operator
\begin{align}
\label{Hoppos}
\hat{H}= \int d^3x \,\,\aopd{\bx} \left (-\frac{\hbar^2}{2 m}\Delta+
  V_{\text{ho}}(\bx)-\mu\right)
\aop{\bx}\nonumber\\
+\frac{1}{2} \int d^6xy \,\,\aopd{\bx} \aopd{\by} \, V_{\text{bin}}(\bx-\by)\,
\aop{\by} \aop{\bx},
\end{align}
where $m$ denotes the atomic mass and $\mu$ is a conveniently chosen
zero-energy reference that will be identified later with the chemical
potential.

In the language of second quantization, the action of a field operator
$\aop{\bx}$ or $\aopd{\bx}$ on a state in Fock space represents the removal or
creation of another unstructured particle at the spatial position $\bx$. The
bosonic nature of the indistinguishable particles is reflected by the
commutation relation of the fields as \be
[\aop{\bx},\aopd{\by}]=\delta(\bx-\by).\ee Due to the symmetry
$[\aop{\bx},\aop{\by}]=0$ of the bosonic particles under coordinate exchange,
only even parity contribution of the interaction potential
$V_{\text{bin}}(\bx)=V_{\text{bin}}(-\bx)$ contribute to the kinematic
evolution.

In principle, all dynamic and static aspects of the evolution of observables
$\hat{\mathcal{O}}$ can be obtained from the solution of Heisenberg's equation
$d\hat{\mathcal{O}}/dt=i\,[\hat{H}, \hat{\mathcal{O}}]/\hbar$ and the
knowledge of the initial state of the system, which is represented by the
many-body density matrix $\brho$. As all observables are formed by the
elementary quantum fields $\aop{}$, it is only necessary to consider the
Heisenberg equation \be
\label{Hosp}
i\hbar \frac{d}{dt}\aop{\bx}(t)=\mathcal{H}(\bx) \, \aop{\bx} + \int d^3y \,
V_{\text{bin}}(\bx-\by)\, \aopd{\by} \aop{\by} \aop{\bx}.\ee For convenience
and later reference, we have introduced in here the Hamilton
operator of a single trapped atom 
\be \label{Hzero} \mathcal{H}(\bx)=-\frac{\hbar^2}{2 m}\Delta+
V_{\text{ho}}(\bx)-\mu.  \ee

In principle, any complete state representation can be used to perform further
calculations. However, it is intuitively clear that a representation that
matches the geometry better or incorporates conserved symmetries of the system
will lead to a greatly simplified description, reveal the essential physics
more clearly and make numerical simulations efficient. Thus, we will decompose
the quantum field in the position representation $\ket{\bx}$ \be
\label{fieldoppos}
\aop{\bx}=\sum_{{i_1,j_1,k_1}} \scal{\bx}{{i_1,j_1,k_1}} \,
\aop{{i_1,j_1,k_1}} \equiv \sum_1 \scal{\bx}{1}\, \aop{1}, \ee in another
complete, yet unspecified basis $\{\ket{1}\equiv \ket{i_1,j_1,k_1}\}$, which
is supposedly more suitable. Furthermore, we will also employ an implicit
summation convention over the quantum labels necessary to specify a state
completely. This means that summation symbols are omitted and a repeated
occurrence of a dummy summation index on one side of an equation implies a
summation.

\begin{fmffile}{nolindiagramsB}
  \fmfcmd{ style_def fsq expr p= cdraw p; cfill (harrow (reverse p,0.55));
    enddef;}

  \fmfcmd{ style_def nsq expr p= cdraw p; cfill (harrow (reverse p,0.5));
    cfill (harrow (p, 0.5)); enddef;}
  \fmfcmd{ style_def msq expr p= cdraw p; cfill (tarrow (reverse p,0.55));
    cfill (tarrow (p, 0.55)); enddef;}
  
  \fmfset{arrow_len}{2.mm} \fmfset{arrow_ang}{20} \fmfset{dash_len}{2mm}
  In this generic basis, the dilute gas Hamiltonian of Eq.~(\ref{Hoppos})
  reads \be
\label{Hopgen}
\hat{H}=\mathcal{H}^{12}\, \aopd{1}\aop{2}+ \phi^{1234}\,
\aopd{1}\aopd{2}\aop{3}\aop{4}, \ee where we have introduced the matrix
elements \bea \mathcal{H}^{12}&=&\bra{1}\mathcal{H}\ket{2}=\int d^3x\,
\scal{1}{\bx}\mathcal{H}(\bx)\scal{\bx}{2}, \eea of the single particle
Hamiltonian Eq.~(\ref{Hosp}) and the two-particle matrix elements \bea
\label{2bdymatelem}
\phiunpr&=&\frac{1}{2}({\mathcal{S}})
\twobra{1}{2}V_{\text{bin}}(\bx_1-\bx_2)\twoket{3}{4}\nonumber\\
&=&\phi^{1243}=\phi^{2134}=\phi^{2143}\nonumber\\
&=&\parbox{1.3cm}{
  \fmfframe(6,16)(6,20){
      \begin{fmfgraph*}(20,20)
        \fmfleft{o1,o2} \fmflabel{1}{o1} \fmflabel{2}{o2}
        \fmfright{i4,i3} \fmflabel{4}{i4} \fmflabel{3}{i3}
        \fmf{plain}{i4,v1,o1} \fmf{plain}{i3,v2,o2}
        \fmf{dashes,tension=0}{v1,v2}
    \end{fmfgraph*}
  }
}, \eea from the binary interaction potential. Due to the bosonic nature of
the particles, only the symmetric part $(\mathcal{S})$ of the instantaneous
coupling vertex $\phiunpr$ is physically relevant and the diagrammatic
representation also carries this property.
\end{fmffile}

In the low kinetic energy range that we are interested in, repulsive s-wave
scattering is the dominant two-particle scattering event
\cite{verhaar694,wieman495}. Provided that a proper T-matrix scattering
calculation has been performed \cite{alkhawaja02,morgan02} or one has obtained
the experimental scattering data, one can encode this information efficiently
via a pseudo potential method \cite{huang,lepage97}.  In the most elementary
invocation of the method, one uses a fictitious contact potential
$V_{\text{bin}}(\bx_1,\bx_2)=V_0\, \delta(\bx_1-\bx_2)$ with a single
parameter $V_0$. This parameter is directly related to the scattering length
$a_{{\rm s}}$ of two particles in vacuo by $V_0=4\pi \hbar^2 a_{{\rm s}}/m$,
provided we would limit all physical approximations to a first order
contribution in $V_0$. Otherwise again an infinite order resummation takes
place and will lead to a renormalization of the effective coupling constant
$V_0$. We will demonstrate this renormalization of the effective coupling
constant takes place in a self-consistent calculation and also leads to a
natural momentum cut-off. In the case of such a contact potential, one finds
for the two-body matrix elements: \bea
\label{deltamatrixel}
\phiunpr&=& \frac{V_0}{2} \int d^3 x\,
\scal{1}{\bx}\scal{2}{\bx}\scal{\bx}{3}\scal{\bx}{4}, \eea which need not be
symmetrized, as it is symmetric already.

In general, no exact solutions of the field equation Eq.~(\ref{Hosp}) or
eigen-states of the Hamiltonian Eq.~(\ref{Hoppos},\ref{Hopgen}) are known and
the few celebrated exceptions \cite{exactsolutions} such as the
Tonks-Girardeau gas \cite{girardeau60,bloch04}, the Lieb-Lininger solution
\cite{shlyapnikov04}, the Richardson pairing model
\cite{richardson68,dukelsky01} and fermionic Luttinger liquids
\cite{schoenhammer96} serve as testing grounds to prove the approximation
schemes.  Fortunately however, most physical phenomena are of universal
character. Thus the system under investigation does not have to follow
precisely a particular model to show a certain response and various
approximations are admissible as long as the main universal aspects of the
problem are incorporated in the specific model hypothesis.

\subsection{Reduced state description with master variables}

The method of quasi-averages and self-consistent field equations has a long
standing tradition in the description of classical gases and fluids
\cite{chapman}, in plasma physics, 
nuclear matter physics \cite{schuck} 
and condensed matter physics
\cite{peletminskii,abrikosov65,blaizot,schrieffer}. It has been
applied successfully to classical particles and degenerate superfluid bosonic
as well as fermionic systems.

The basic premises for a reduced state description of a weakly correlated
many-body problem relies on the existence of a well separated hierarchy of
time, energy and length scales. If this is the case, one can assume that the
information required to describe an ensemble effectively can be parameterized
with a set of a few relevant variables \cite{peletminskii}. However
the price that has to be payed for reducing the ''astronomical'' dimension of
the linear many-body Schr\"odinger equation is giving up the superposition
principle and embarking on non-linear mathematics.

In the mean-field approximation, we want to assume that the ensemble of
relevant quantum states establishes a well defined mean value for the field
operator and that quantum fluctuations only cause small deviations
around it \bea 
\label{symbreak}
\aop{1}&=&\av{\aop{1}}+\delta \aop{1},\eea such that $\av{\delta \aop{1}}=0$.
As mentioned before in Eq.~(\ref{fieldoppos}), we use in here the shorthand
notation for any complete set quantum of labels e.g., $1\equiv \bx_1$ in
position space or $1\equiv \boldmath{k}_1$ in momentum space, respectively.
This number symmetry breaking approximation is tremendously useful and can be
envisaged also as the semi-classical limit of coherent many-particle quantum
physics in analogy to the description of the optical laser \cite{schleich01}.
However, one must be aware of the implied consequences that have been
discussed in the literature
\cite{Hohenberg1965,griffin496,Stoof1999a,leggett401} and number conserving
approximations \cite{girardeau59,Castin1998a,zollerv} have their merits, but
shortcomings as well.

\begin{fmffile}{nolindiagramsA}
  \fmfcmd{ style_def fsq expr p= cdraw p; cfill (harrow (reverse p,0.55));
    enddef;}

  \fmfcmd{ style_def nsq expr p= cdraw p; cfill (harrow (reverse p,0.5));
    cfill (harrow (p, 0.5)); enddef;}

  \fmfcmd{ style_def msq expr p= cdraw p; cfill (tarrow (reverse p,0.55));
    cfill (tarrow (p, 0.55)); enddef;}
  
  \fmfset{arrow_len}{2.mm} \fmfset{arrow_ang}{20} \fmfset{dash_len}{2mm}
  Based on these assumptions, we want to sum the c-number field amplitudes
  $\al{1}=\av{\aop{1}}$ that were introduced in Eq.~(\ref{symbreak}) over this
  complete set of states and form a basis-independent element of a Hilbert
  space as \bea \ket{\alpha}= &\al{1}\,\ket{1} \equiv
  \av{{\hat{a}}}=\parbox{1.1cm}{ \fmfframe(3,6)(6,6){
    \begin{fmfgraph*}(24,12)                                
      \fmfleft{o2}
      \fmfright{i1}
      \fmf{wiggly}{i1,o2}
    \end{fmfgraph*}
  } }.\eea In here, we have used the Dirac notation with the implied standard
scalar product $\scal{\alpha}{\beta}=\int d^3x\, \alpha^\ast(\bx)\beta(\bx)$.
Due to the non-linear nature of the ensuing mean field equations, we can no
longer rely on the dynamical superposition principle available in linear
quantum mechanics. However, the advantage of the notation arises from
preserving and emphasizing the geometrical transformations properties of all
correlations functions under a change of basis or, most generally, frame of
reference.  The wiggly line has been introduced to represents the mean-field
amplitude graphically. This symbol literally denotes the state at time $t$ and
not the time-ordered propagator that evolves it.

In an analogous fashion, we can separate the single particle density operator
of the atomic gas $\ft{}=\av{ {\hat{a}}^{\dag} \hat{a} }=\fc{}+\fs{}$ into a
mean-field contribution $\fc{}$ and a fluctuation $\fs{}$ around it.  This
contribution $\fs{}$ is also known as the normal quantum depletion of the
atomic cloud. Both quantities are hermitian tensor operators of rank (1,1) and
defined as
\begin{align}
  \label{fcfsq}      
\fc{}=\ket{\alpha} \bra{\alpha},\quad \fs{}=\fs{14}\,\ket{1}\bra{4}=
\parbox{1.1cm}{ \fmfframe(3,6)(12,6){
    \begin{fmfgraph*}(24,12)
      \fmfleft{o1}
      \fmfright{i4}
      \fmf{plain_arrow}{i4,o1}
    \end{fmfgraph*}
  } }.
\end{align}
For example, in a position representation, the single particle density matrix
reads as
$\bra{\bx_1}f\ket{\bx_2}=\fc{}(\bx_1,\bx_2)+\fs{}(\bx_1,\bx_2)=
\alpha(\bx_1)\alpha^\ast(\bx_2)+\av{(\aopd{\bx_2}-\alpha^\ast(\bx_2))
  (\aop{\bx_1}-\alpha(\bx_1))}$

Similarly, we define anomalous averages or pairing fields $\mt{}=\av{\hat{a}
  \hat{a}}=\mc{}+\ms{}$, as symmetric tensors of rank (2,0), \ie, \bea
\label{mcmsq}       
\mc{}=\ket{\al{}}\ket{\al{}},\quad
\ms{}=\ms{12}\,\ket{1}\ket{2}=\parbox{1.1cm}{ \fmfframe(3,6)(12,6){
    \begin{fmfgraph*}(24,12)
      \fmfleft{o1}
      \fmfright{i2}
      \fmf{msq}{i2,o1}
    \end{fmfgraph*}
  } }, \eea as well as  their symmetric conjugates as $
\ms{}^\ast=
\parbox{1.1cm}{
  \fmfframe(0,0)(0,0){
    \begin{fmfgraph*}(24,12)
      \fmfleft{o1}
      \fmfright{i2}
      \fmf{nsq}{i2,o1}
    \end{fmfgraph*}
  } }$. These anomalous averages are also associated with the quantum
depletion of the degenerate gas. In contrast to the single particle nature of
normal density matrix of Eq.~(\ref{fcfsq}), it can be seen immediately from
the structure of Eq.~(\ref{mcmsq}) that the pairing field is a two-particle
state. It will be shown in Sec.~\ref{secTmat} that the pairing field
$\ms{}(\bx_1,\bx_2)$ evolves basically like a generalization the bare
two-particle Schr\"odinger state. Thus, it carries all the important physics
of binary scattering.  In a position representation, the pairing field
is given by $\bra{\bx_1}\bra{\bx_2}\,m=\mc{}(\bx_1,\bx_2)+\ms{}(\bx_1,\bx_2)=
\alpha(\bx_1)\,\alpha (\bx_2)+\av{(\aop{\bx_2}-\alpha(\bx_2))
  (\aop{\bx_1}-\alpha(\bx_1))}$.
\end{fmffile}

\subsection{Structure of the generalized density matrix}
\label{secG}

From the transformation properties of the matrices under coordinate change,
one finds that the normal density matrix $\fs{}(t)$ and the pairing fields
$\ms{}(t)$ are not independent but actually the components of a generalized
density matrix $G$ \cite{kadanoff62,schuck,blaizot}. If we represent the
system in a state space of dimension $n$, then it is convenient to arrange the
mean field in a 2n-dimensional row vector $\chi$ and the fluctuations as a
positive semi-definite matrix $\text{dim}[G]=2n\times 2n$,
\begin{align}
\label{chi}
  \chi=\left(
\begin{array}{c}
  \al{}\\
  \ald{}
\end{array}
\right), \quad
G= \left(
  \begin{array}{cc}
    \fs{} & \ms{}\\
    \ms{}^\ast & \fspo{}^\ast
  \end{array}
\right)\ge0.
\end{align}
The non-negativity of this co-variance matrix implies that the magnitude of
the anomalous fluctuations is limited by the normal depletion through a
Cauchy-Schwartz inequality (see Appendix \ref{CauchySchwartz}).  In the
general context of Green function's, this single-time density operator $G(t)$
can also be viewed as a particular limit of a time-ordered (${\cal T}$),
two-time Green function $G(\tau,t)$, \ie, $G(t)\equiv \Gg(t)=\lim_{\tau
  \rightarrow t_+} {\cal T} \,G(\tau,t)$. Consequently, it is also necessary
to consider the opposite limit and to define a time-reversed, single-time
density operator through $\Gk(t)=\lim_{\tau\rightarrow t_-} {\cal T}\,
G(\tau,t)$.  Explicitly, this operator is given by \bea
\label{Gk}
\Gk&=&\paul{1}\, {\Gg}^\ast \paul{1}=\Gg+\paul{3}=
\left(
  \begin{array}{cc}
    \mathds{1}+\fs{}& \ms{}\\
    \ms{}^\ast& \fs{}^\ast
  \end{array}
\right), \eea where standard Pauli spin matrices have been introduced and are
defined in Appendix \ref{appa}.

The specific structure of the generalized density matrix implies various
important physical properties. First of all, we have to assume that there is a
basis that diagonalizes this fluctuation matrix.  Exactly $n$ of its $2\,n$
eigen-values correspond to the positive occupation numbers of finding a
particle or, more generally, a quasi-particle in a certain mode.  For a given,
but otherwise arbitrary, $G$ matrix, one can construct this basis by studying
the transformation law of the density matrix under a canonical transformation
$T$ (see Appendix \ref{canonical}), \bea G^\prime=T\, G\, T^\dag. \eea It is
important to note that this is not the transformation law of a general matrix
under coordinate change. This would require that $T^\dag=T^{-1}$.  However, by
only using the properties of the symplectic transformations, one can show that
a canonical eigen-value problem is defined by \bea \left( \paul{3}\,G \right)
\,T^\dag &=&T^\dag \, \left( \paul{3}\, G^\prime \right). \eea The solution of
this eigen-value problem yields the eigen-vector matrix $T^\dag$ and the
corresponding diagonal eigen-value matrix $\paul{3}\,G^\prime$.  All
normalizable states can be rescaled such that $T\paul{3}\,T^\dag=\paul{3}$.
Now, we are able to reconstruct the positive $G$ matrix \bea \label{specdecG}
G&=&V\,P\,V^\dag,\eea from its eigen-vectors $V=\paul{3}\,T^\dag$ and the
diagonal, positive occupation number matrix $P=\paul{3}\,G^\prime \paul{3}$.

Second, an important feature of an admissible fluctuation matrix is its
consistency with the commutation relation, \ie,
$\av{\aop{1}\aopd{2}}=\av{\aopd{2}\aop{1}}+\delta_{12}$ and
$\av{\aop{1}\aop{2}}=\av{\aop{2}\aop{1}}$.  This was already expressed in
Eq.~(\ref{Gk}) as \bea \paul{1}\,G^\ast\,\paul{1} - G &=& \paul{3}.  \eea By
invoking the properties of a unitary symplectic transformation, one can show
that the elements of the diagonal occupation number matrix $P$ are not $2\,n$
independent variables. Actually half of them are determined by the other half,
$P_{(n+1,\dots, 2\,n)}=1+P_{(1,\dots,n)}$, or \bea \paul{1} \, P\,
\paul{1}-P=\paul{3}.  \eea In other words, by separating the occupation
numbers $P$ and the eigen-vector matrix $V$ into a first and second half, \ie,
$P=(P_{+},\mathds{1}+P_{+})$ and $V=(V_{+},V_{-})$, one can then decompose a
general fluctuation matrix as \bea
\label{genericgmat}
G&=V_{+}^{\phantom{\dag}}\, P_{+}^{\phantom{\dag}} \, V_{+}^\dag
+V_{-}^{\phantom{\dag}}\,(\mathds{1}+P_{+}^{\phantom{\dag}})\,V_{-}^\dag. \eea

\subsection{Dynamic equations of motion}
\label{secDyn}
It is now straight forward to derive equations of motion for these averages
directly from the Heisenberg equation Eq.~(\ref{Hosp}). However, due to the
nonlinear character of the operator equation, one finds always a coupling of
correlation functions involving $n$ fields to a correlation function of $n+2$
fields. If we truncate this infinite hierarchy of correlation functions
(BBGKY) at the level of one and two-operator fields $\{\al{}, \fs{}, \ms{}\}$
and approximate higher correlation functions with the help of Wick's theorem
\cite{peletminskii,walser599,wachter501}, one obtains the following equations
of motion, also known as time-dependent Hartree-Fock Bogoliubov (THFB)
equations (THFB) \bea
\label{collisionlesso}
i\hbar \frac{d}{dt}\chi&=&\Pi\,\chi,\\
\label{collisionlesst}
i\hbar \frac{d}{dt}G&=&\Sigma\,G-G \,\Sigma^\dag.\eea

For the evolution of the mean-field $\chi$, one finds a generalized
Gross-Pitaevskii propagator that is defined as 
\bea 
\label{piGP}
\piGP{}&=& \left(
  \begin{array}{cc}
    \piGP{N}& \piGP{A}\\
    -\piGP{A}^\ast & -\piGP{N}^\ast
  \end{array}
\right). \eea The two contributions that define this symplectic propagator are
a normal hermitian Hamiltonian operator and a symmetric anomalous coupling
potential
\begin{fmffile}{nolindiagramsC}
  \fmfcmd{ style_def fsq expr p= cdraw p; cfill (harrow (reverse p,0.55));
    enddef;}

  \fmfcmd{ style_def nsq expr p= cdraw p; cfill (harrow (reverse p,0.5));
    cfill (harrow (p, 0.5)); enddef;}
  \fmfcmd{ style_def msq expr p= cdraw p; cfill (tarrow (reverse p,0.55));
    cfill (tarrow (p, 0.55)); enddef;}
  
  \fmfset{arrow_len}{2.mm} \fmfset{arrow_ang}{20} \fmfset{dash_len}{2mm}
\begin{align}
\label{Hc}
\piGP{N}&=\mathcal{H}+\Uc+2\,\Usq\\\nonumber
&= \parbox{0.7cm}{
  \fmfframe(6,6)(0,6){
     \begin{fmfgraph*}(12,24)
       \fmfleft{o1}
       \fmfright{i2}
       \fmf{plain}{i2,o1}
     \end{fmfgraph*}
   }
 } + \parbox{1.4cm}{
   \fmfframe(6,0)(0,6){
      \begin{fmfgraph*}(30,20)
        \fmfleft{o1,o2}
        \fmfright{i4,i3}
        \fmftop{t1,t2}
        \fmf{plain}{i4,v1,o1} \fmf{wiggly}{t1,v2,t2}
        \fmf{dashes,tension=0}{v1,v2}
      \end{fmfgraph*}
    }
  } +2\times \parbox{1.4cm}{
    \fmfframe(0,24)(0,20){
        \begin{fmfgraph*}(24,14)      
          \fmfleft{o1,o2}
          \fmfright{i4,i3}
          \fmftop{t}
          \fmf{plain}{i4,v1,o1}
          \fmf{dashes,tension=0}{v1,t}
          \fmf{fsq,tension=0.7}{t,t}
      \end{fmfgraph*}
    }
  },\\
\label{Oc}
\piGP{A}&=\Vsq= \parbox{1.6cm}{
  \fmfframe(6,10)(0,12){
    \begin{fmfgraph*}(30,24)
      \fmfleft{o1,o2}
      \fmfright{i4,i3}
      \fmf{phantom,tension=1}{i3,v2} \fmf{phantom,tension=1}{i4,v1}
      \fmf{plain}{v2,o2} \fmf{plain}{v1,o1}
      \fmf{dashes,tension=0.0}{v1,v2}
      \fmf{msq,right=1,tension=0.0}{v1,v2}
    \end{fmfgraph*}
  }
}.
\end{align} 
It is easy to identify $\piGP{N}$ with the well known hermitian GP-propagator
that accounts for the free evolution of the mean-field $\mathcal{H}$, its
self-interaction $\Uc$, as well as the energy shift $\Usq$, which caused by
the presence of the non-condensate cloud.  However, due to the existence of
the anomalous fluctuations there is also a coupling through $\piGP{A}$ to the
time-reversed field.  For convenience, we have introduced two auxiliary
potentials $\mathcal{U}_{\ft{}}$ and $\mathcal{V}_{\mt{}}$.  Explicitly, they
are defined in terms of the two-body matrix elements as
\begin{align} 
  \mathcal{U}_{\ft{}}^{14}&= 2\,\phi^{1234}\,\ft{3 2}, \\
  \mathcal{V}_{\mt{}}^{12}&=2\,\phi^{1234}\,\mt{3 4}.
\end{align}
In a position representation, this reduces 
to the familiar non-local Hartree-Fock potentials
\begin{align} 
  \mathcal{U}_{\ft{}}(\bx,\by)=&
  \frac{1}{2} [ V_{\text{bin}}(\bx-\by)\,\ft{}(\bx,\by)\\
  +&\delta(\bx-\by)\,\int d^3z \,V_{\text{bin}}(\bz-\by)\,
  \ft{}(\bz,\bz) ], \nonumber\\
  \mathcal{V}_{\mt{}}(\bx,\by)=&V_{\text{bin}}(\bx-\by)\,\mt{}(\bx,\by).
\end{align}

Similarly, one finds that the evolution of the density operator $G$ is ruled
by a HFB self-energy  $\Sigma$ that can be
obtained also by variational methods \cite{blaizot}. In detail, this
symplectic self-energy is given by 
\bea 
\label{sigmaHFB}
\Sigma&=& \left(
  \begin{array}{cc}
    \sigN& \sigA\\
    -\sigA^{\ast} & -\sigN^\ast
  \end{array}
\right), \eea where we have introduced hermitian Hamiltonian operators and
symmetric anomalous coupling potentials as \bea
\label{Hsq}
\sigN&=&\mathcal{H}+2\,\Uc+2\,\Usq\\
&=& \parbox{0.8cm}{
  \fmfframe(6,6)(0,6){
     \begin{fmfgraph*}(12,24)
       \fmfleft{o1} 
       \fmfright{i2} 
       \fmf{plain}{i2,o1}
     \end{fmfgraph*}
   }
 } + 2\times \parbox{1.4cm}{
   \fmfframe(0,0)(0,6){
      \begin{fmfgraph*}(30,20)
        \fmfleft{o1,o2} 
        \fmfright{i4,i3} 
        \fmftop{t1,t2}
        \fmf{plain}{i4,v1,o1} 
        \fmf{wiggly}{t1,v2,t2}
        \fmf{dashes,tension=0}{v1,v2}
      \end{fmfgraph*}
    }
  } +2\times \parbox{1.0cm}{
    \fmfframe(0,24)(0,20){
        \begin{fmfgraph*}(24,14)      
          \fmfleft{o1,o2} 
          \fmfright{i4,i3} 
          \fmftop{t}
          \fmf{plain}{i4,v1,o1}
          \fmf{dashes,tension=0}{v1,t}
          \fmf{fsq,tension=0.7}{t,t}
      \end{fmfgraph*}
    }
  }.\nonumber\\
\label{Osq}
\sigA&=&\Vc+\Vsq= \parbox{1.5cm}{
  \fmfframe(6,10)(0,12){
      \begin{fmfgraph*}(30,24)
        \fmfleft{o1,o2} 
        \fmfright{i4,i3}
        \fmf{plain}{o1,v1} \fmf{wiggly}{v1,i4} \fmf{plain}{o2,v2}
        \fmf{wiggly}{v2,i3}
        \fmf{dashes,tension=0}{v1,v2}
      \end{fmfgraph*}
    }
  } + \parbox{1.5cm}{
    \fmfframe(6,10)(0,12){
      \begin{fmfgraph*}(30,24)
        \fmfleft{o1,o2} 
        \fmfright{i4,i3}
        \fmf{phantom,tension=1}{i3,v2} \fmf{phantom,tension=1}{i4,v1}
        \fmf{plain}{v2,o2} \fmf{plain}{v1,o1}
        \fmf{dashes,tension=0.0}{v1,v2}
        \fmf{msq,right=1,tension=0.0}{v1,v2}
      \end{fmfgraph*}
    }
  }.  \eea
\end{fmffile}
It is important to note the different weighing factors of the mean-field
potential in Eqs.~(\ref{Hc}) and (\ref{Hsq}) and the fact that the potentials
are local in time.

\subsection{Structure of the Hartree-Fock-Bogoliubov self energy}
\label{secHFB}
\subsubsection{Normal quasi-particle modes}
Symplectic self-energy operators arises not only naturally in kinetic theories
\cite{kadanoff65,wachter501} or variational calculations, but in many other
contexts involving stability analysis.  In the case of bosonic fields, the
self-energy operator is of the generic form: \bea \Sigma&=& \left(
  \begin{array}{cc}
    \sigN & \sigA\\
    -{\sigA}^\ast & -{\sigN}^\ast
  \end{array} 
\right).  \eea In here, $\sigN$ stands for a hermitian operator
$\sigN=\sigN^\dag$ and $\sigA$ denotes an anomalous coupling term that has to
be symmetric $\sigA=\sigA^\top$.  The relative size of the operators $\sigN$
and $\sigA$ determines the character of the energy spectrum. It can either be
real valued with pairs of positive and negative eigen-energies, or one finds a
doubly degenerate zero eigen-value, if the energy difference between the
smallest positive and highest negative vanishes (gap-less spectrum).  In the
general case, there is a mixed spectrum consisting of pairs of real
sign-reversed as well as pairs of complex conjugated eigen-values. The
eigen-vectors $W$ are normalizable with respect to the indefinite norm
$|\!|W|\!|^2=W^\dag\paul{3}W$, except for those that belong to zero or complex
eigen-values.  It is important to note that this energy basis $W$ is in
general distinct from the instantaneous basis $V$ that diagonalizes the
fluctuation matrix $G$ in Eq.~(\ref{specdecG}). They do coincide only in
equilibrium.  The mathematical properties of the eigen-states $W$ can be
derived easily from the intrinsic symmetries of the HFB self-energy operator:
\bea
\label{sym1}
\Sigma&=&-\paul{1}\,\Sigma^\ast \paul{1},\\
\label{sym2}
\Sigma^\dag&=&\paul{3}\,\Sigma \,\paul{3}.  \eea Thus, if $W$ is a solution of
the right eigen-value problem with energy $E$, \bea
\label{rightev}
\Sigma\,W=W\,E, \eea 
it follows directly from Eq.~(\ref{sym1}) that
$\overline{W}=\paul{1}\,W^\ast$, is also a right eigen-vector but
corresponds to the eigen-value $\bar{E}=-E^\ast$. 
Starting from the second symmetry in Eq.~(\ref{sym2}) and the right
eigen-value problem of Eq.~(\ref{rightev}), it is easy to construct
the left eigen-vectors $\widetilde{W}=W^\dag\, \paul{3}$ that correspond
to the eigen-values $\tilde{E}=E^\ast$: \bea \widetilde{W}\,
\Sigma=E^\ast\,\widetilde{W}.  \eea
Finally, from a combination of the results for the right and left
eigen-vectors, it follows that the eigen-vectors are orthogonal with
respect to the metric $\paul{3}$: \bea
0=(E^\ast-E^\prime)\,W_{E}^\dag\paul{3}
W_{{E^\prime}}^{\phantom{\dag}}, \eea if $E^\ast \neq E^\prime$. On
the other hand, this relation implies also that eigen-vectors that
belong to complex eigen-values must have zero norm.

\subsubsection{Defective sub-spaces}
The situation of a doubly degenerate zero energy eigen-value $E=0$ needs
special attention.  One can view this case as a limit when two non-degenerate
states approach each other. However, as the energy gap decreases, the two
eigen-states become more and more collinear. Thus, in the limit of a vanishing
energy separation, the dimension of the spanned vector space collapses from 2
to 1 and $\Sigma$ becomes defective \cite{blaizot,golub96}.  In the present
context, a gap-less linear response matrix occurs from a perturbation analysis
of the simple Gross-Pitaevskii equation and describes the collective
excitation of the system. The emerging zero energy state is called Goldstone
mode and can be interpreted physically as an attempt to restore the broken
number symmetry. On the other hand, we find for the THFB
Eqs.~(\ref{collisionlesso},\ref{collisionlesst}) that the self-energy matrix
$\Sigma$ has a gap and its eigen-states form a complete non-defective basis.
It is important to distinguish these states from the collective excitations of
the total system, whose excitation spectrum is again gap-less
\cite{giorgini00,morgan04}.

The general situation can be described by separating the two ground state
modes $(W_{E_0},W_{-E_0})$ from the remaining states $W^\prime$ such that
dim[$W^\prime$]=$2n\times 2(n-1)$ and the diagonal eigenvalue matrix has a
dim[$E^\prime$]=$2(n-1)\times 2(n-1)$.  By introducing two quadrature modes
$PQ$ with dim[$PQ$]=$2n\times 2$ via \bea
\Sigma\,W^\prime&=&W^\prime E^\prime,\\
\Sigma\, PQ&=&PQ\, i\hbar\,\left(
  \begin{array}{cc}
    0& -M^{-1}\\
    M \omega_0^2 & 0
  \end{array}
\right), \eea one can span a two dimensional vector space which is orthogonal
to the higher modes $W^\prime$ and does not collapse
\cite{Lewenstein1996a,You1998}.  This is mathematically achieved by the
construction of a ``best'' basis in the context a singular value decomposition
\cite{golub96} and introduces two singular values, \ie, an inertial mass
parameter $M$ and a gap energy $E_0=\hbar\, \omega_0$ of the number and phase
quadratures.  The quadrature states satisfy the following orthogonality
\begin{gather}
PQ^\dag \paul{3} W^\prime=0,\quad
PQ^\dag \paul{3} PQ=\hbar\, \paul{2}^{(1)},
\end{gather}
and symmetry relations $PQ= -\paul{1} PQ^\ast$. From a dimensional
consideration, it is obvious that one also has to use lower dimensional Pauli
matrices, which act in the appropriate subspaces such that
dim[$\paul{k=1,2,3}^{(l)}$]=$2l\times 2l$.  All states together form again a
complete basis such that
\begin{gather}
\label{completeness}
W^\prime \paul{3}^{(n-1)} {W^\prime}^\dag+PQ \,\frac{\paul{2}^{(1)}}{\hbar} 
\,PQ^\dag=\paul{3}.
\end{gather}
With these states, we can then obtain the following spectral decomposition of
the self-energy
\begin{gather}
  \Sigma\paul{3}=W^\prime(E^\prime\paul{3}^{(n-1)}){W^\prime}^\dag+
  PQ \left(
    \begin{array}{cc}
      M^{-1} &0\\
      0 & M \omega_0^2
    \end{array}
  \right) PQ^\dag\nonumber\\
  =W^\prime(E^\prime\paul{3}^{(n-1)}){W^\prime}^\dag+
  \frac{1}{M}\,P\cdot P^\dag+M\omega_0^2\,Q\cdot Q^\dag.
\end{gather}
The physical meaning of the quadrature states can be understood most clearly
when mapping them again back onto the quantum field 
\begin{align}
 \left(
\begin{array}{c}
  \hat{P}\\
  \hat{Q}     
\end{array} \right)
&=PQ^\dag \sigma_3 \left(
  \begin{array}{c}
    \aop{}\\
      \aopd{}
    \end{array}
  \right),\,
 \left(
\begin{array}{c}
  \hat{b}^{\phantom \dag}\\
  \hat{b}^\dag     
\end{array} \right)
={W^\prime}^\dag \sigma_3 \left(
  \begin{array}{c}
    \aop{}\\
    \aopd{}
  \end{array}
\right).
\end{align}
Now, these field quadratures $\hat{Q}$, $\hat{P}$ satisfy a conventional
Heisenberg-Weyl algebra and define in depend, as well as orthogonal bosonic
quasi-particles  $\hat{b}_i$ according to 
\begin{gather}
\label{quasiparticles}
 \comut{\hat{Q}}{\hat{P}}=i\,\hbar,\quad
 \comut{\hat{b}^{\phantom\dag}_i}{\hat{b}^\dag_j}=\delta_{ij}.
\end{gather}
  
\subsection{Upgrading off-diagonal potentials to many-body T-matrices}
\label{secTmat}
\begin{fmffile}{fragA}
  \fmfcmd{ style_def fsq expr p= cdraw p; cfill (harrow (reverse p,0.55));
    enddef;}
  
  \fmfcmd{ style_def nsq expr p= cdraw p; cfill (harrow (reverse p,0.5));
    cfill (harrow (p, 0.5)); enddef;}
  \fmfcmd{ style_def msq expr p= cdraw p; cfill (tarrow (reverse p,0.55));
    cfill (tarrow (p, 0.55)); enddef;}
  
  \fmfset{arrow_len}{2.mm} \fmfset{arrow_ang}{20} \fmfset{dash_len}{2mm}
  While one can understand the physical structure of the self-energy best in
  the general form of the kinetic equations Eq.~(\ref{collisionlesst}), one
  can appreciate other aspects much better by considering the equations for the
  components individually. 
  \begin{gather}
    i\hbar\frac{d}{dt} \fs{}=\sigN \,\fs{} -\fs{} \,\sigN
    +\sigA\, \ms{}^\ast -\ms{}\,\sigA^\ast, \\
    i\hbar\frac{d}{dt} \ms{}=\sigN \,\ms{} +\ms{} \,\sigN^\ast
    +\sigA\, \fspoT{} + \fs{}\,\sigA, 
  \end{gather}
  In stationarity and by assuming a real valued self-energy, one can solve for
  $\ms{}$ in an HF eigen basis 
  \be
  \label{sigHF}
  \sigN\;\ket{\epsilon_i}=(\epsilon_i-\mu)\ket{\epsilon_i}, \ee and finds \be
  \label{msstat}
  \ms{12}=\frac{\sigA^{13}\,\fspoT{32}
    +\fs{13}\,\sigA^{32}}{2\mu-(\epsilon_1+\epsilon_2)}.  \ee No rules for
  circumventing poles in the above energy denominator need to be specified as
  all energies satisfy $\epsilon_i>\mu$.

In the present article, we have only considered first order processes 
and found the one-loop contribution to the self-energy as 
\begin{gather}
  \sigA=\Vc+\Vsq= \parbox{1.5cm}{
    \fmfframe(6,10)(0,12){
      \begin{fmfgraph*}(30,24)
        \fmfleft{o1,o2} 
        \fmfright{i4,i3}
        \fmf{plain}{o1,v1} \fmf{wiggly}{v1,i4} \fmf{plain}{o2,v2}
        \fmf{wiggly}{v2,i3}
        \fmf{dashes,tension=0}{v1,v2}
      \end{fmfgraph*}
    }
  } + \parbox{1.5cm}{
    \fmfframe(6,10)(0,12){
      \begin{fmfgraph*}(30,24)
        \fmfleft{o1,o2} 
        \fmfright{i4,i3}
        \fmf{phantom,tension=1}{i3,v2} \fmf{phantom,tension=1}{i4,v1}
        \fmf{plain}{v2,o2} \fmf{plain}{v1,o1}
        \fmf{dashes,tension=0.0}{v1,v2}
        \fmf{msq,right=1,tension=0.0}{v1,v2}
      \end{fmfgraph*}
    }
  }.  \end{gather} Now, if we substitute the stationary solution of
Eq.~(\ref{msstat}) into this equation, one obtains an inhomogeneous linear
equation for the self-energy \be \sigA^{12}=2\phiunpr\mc{34} +2\phiunpr
\frac{(\mathds{1}+2\fs{})_{3 3^\prime}}{2\mu-(\epsilon_{3} +\epsilon_{4})}
\sigA^{3^\prime 4}.  \ee It is now clear that the value of $\sigA$ is only
determined by $\mc{}$ in the absence of a homogeneous contribution.  Thus,
there must be a T-matrix that maps mean-field to the off-diagonal self-energy
\be 
\label{TA}
\sigA^{12}=T_{\cal A}^{1234}(2\mu)\,\mc{34}= \parbox{1.9cm}{
    \fmfframe(12,6)(0,6){
    \begin{fmfgraph*}(36,24)
      \fmfstraight
      \fmfleftn{l}{2} \fmfrightn{r}{2}
      \fmfrpolyn{filled=20,tension=0.5}{k}{4} \fmf{plain}{l2,k4}
      \fmflabel{2}{l2} \fmf{plain}{l1,k3}
      \fmflabel{1}{l1}                %
      \fmf{phantom}{k4,k1} \fmf{phantom}{k3,k2}
      \fmf{wiggly}{k1,r2} \fmf{wiggly}{k2,r1}
                                %
    \end{fmfgraph*}
  }
}. \ee This leads to the following general definition for an off-the energy
shell T-matrix \be T_{\cal A}^{1234}(\epsilon)=2\phiunpr+2\phi^{123^\prime
  4^\prime} \frac{(\mathds{1}+2\fs{})_{3^\prime
    3^\ppr}}{\epsilon-(\epsilon_{3^\prime} +\epsilon_{4^\prime})} T_{\cal
  A}^{3^\ppr 4^\prime 34}.  \ee In operator notation this is equivalent to \be
T_{\cal A}(\epsilon)=V+V G_{\cal A}(\epsilon) [(\mathds{1}+2\fs{})\otimes
{\mathds{1}}] T_{\cal A}(\epsilon).  \ee where $G_{\cal
  A}(\epsilon)=(\epsilon-\sigN \otimes{\mathds{1}}-{\mathds{1}} \otimes
\sigN)^{-1}$ is a Green's function, which describes the propagation of two
independent HF particles according to Eq.~(\ref{sigHF}).  The diagrammatic
representation of this $T_{\cal A}$-matrix relation is

\be \parbox{1.4cm}{
  \fmfframe(0,0)(0,0){
    \begin{fmfgraph*}(36,24)
      \fmfstraight
      \fmfleftn{l}{2} \fmfrightn{r}{2}
      \fmfrpolyn{thick,filled=20,tension=0.5}{k}{4} \fmf{plain}{l2,k4}
      \fmf{plain}{l1,k3}               %
      \fmf{phantom}{k4,k1} \fmf{phantom}{k3,k2}
      \fmf{plain}{k1,r2} \fmf{plain}{k2,r1}
                                %
    \end{fmfgraph*}
  }
}= \parbox{1.4cm}{
  \fmfframe(0,0)(0,0){
    \begin{fmfgraph*}(36,24)
      \fmfstraight
      \fmfleftn{l}{2} \fmfrightn{r}{2} \fmf{plain,tension=3}{l2,v2}
      \fmf{plain,tension=3}{l1,v1}
                %
      \fmf{dashes,tension=2}{v1,v2} \fmf{plain,tension=3}{v1,r1}
      \fmf{plain,tension=3}{v2,r2}
    \end{fmfgraph*}
  }
}+ \parbox{1.4cm}{
  \fmfframe(0,0)(0,0){
    \begin{fmfgraph*}(36,24)
      \fmfstraight
      \fmfleftn{l}{2} \fmfrightn{r}{2} \fmf{plain,tension=3}{l2,v2}
      \fmf{plain,tension=3}{l1,v1}
                %
      \fmf{dashes,tension=1.7}{v1,v2} \fmfrpolyn{thick,filled=20}{k}{4}
      \fmf{plain_arrow}{v1,k3} \fmf{plain}{v2,k4} \fmf{plain,tension=3}{k2,r1}
      \fmf{plain,tension=3}{k1,r2}
    \end{fmfgraph*}
  }
}+ \parbox{1.4cm}{
  \fmfframe(0,0)(0,0){
    \begin{fmfgraph*}(36,24)
      \fmfstraight
      \fmfleftn{l}{2} \fmfrightn{r}{2} \fmf{plain,tension=3}{l2,v2}
      \fmf{plain,tension=3}{l1,v1}
                %
      \fmf{dashes,tension=1.7}{v1,v2}
      \fmfrpolyn{width=thick,filled=20,tension=1}{k}{4}
      \fmf{plain,tension=1}{v1,k3} \fmfpen{thick}
      \fmf{plain_arrow,tension=1}{k4,v2} \fmfpen{thin}
      \fmf{plain,tension=3}{k2,r1} \fmf{plain,tension=3}{k1,r2}
    \end{fmfgraph*}
  }
}. \ee
\end{fmffile}

\subsection{Invariants and Conservation laws}

\subsubsection{Number}
The total particle number $\hat{N}=\int d^3 x \,\aopd{\bx}\aop{\bx}$ is a
conserved quantity if the atoms evolve under the generic two-particle
Hamiltonian operator $\hat{H}$ given by Eq.~(\ref{Hoppos}), \ie,
$[\hat{H},\hat{N}]=0$.  This conservation law implies that the system is
invariant under a global phase change $\aop{}\rightarrow
\aop{}\,\exp{(-i\,\Phi)}$.  By using this continuous symmetry, \ie,
$\al{}\rightarrow \al{}\,\exp{(-i\,\Phi)}$, $\fs{}\rightarrow \fs{}$, and
$\ms{}\rightarrow \ms{}\, \exp{(-2\,i\,\Phi)}$, it is easy to see that the
kinetic equations Eqs.~(\ref{collisionlesso},\ref{collisionlesst}) are also
explicitly number conserving at all times: \bea
N(t)=\av{\hat{N}}&=&\text{Tr}\left\{\fc{}(t)+\fs{}(t)\right\}=\text{const.}
\eea Nevertheless, it is important to note that out-of equilibrium there can
be a continuous particle exchange between the condensate and the
non-condensate clouds.

\subsubsection{Energy}
In the absence of any time-dependent external driving fields, such as optical
lasers or magnetic rf-fields, the overall energy $\hat{H}$ must be conserved
as well.  To find the expectation value of the total system energy
$E=\av{\hat{H}}$, we use again Wick's theorem (see Appendix~\ref{wick})
systematically.  Explicitly, this energy functional is given as
\begin{gather}
  E(t)=\text{Tr} \left\{ [ \mathcal{H}+\frac{1}{2}\Uc+\Usq]\, \fc{}
    + [ \mathcal{H}+\Uc+\Usq] \,\fs{} \right.\nonumber\\
  +\left.  \frac{1}{2}\, \Vsq{}\,{\mc{}}^\ast+ \frac{1}{2}\,[\Vc{}+\Vsq{}]
    \,\ms{}^\ast\right\}=\text{const.}
\end{gather}
That this is also a constant of motion follows straight from 
Eqs.~(\ref{collisionlesso},\ref{collisionlesst}) and make this a 
``derivable theory'' according to Ref.~\cite{Hohenberg1965}.
For example, the same first order results can be found in Ref.~\cite{blaizot},
derived by a variational procedure.

\subsubsection{Entropy and individual occupation probabilities}

In a previous section Sec.~\ref{secG}, we have demonstrated with
Eq.~(\ref{specdecG}) that any admissible covariance matrix is necessarily of
the following form 
\begin{gather}
\label{DefGg}
G(t)=V(t)\,P(t)\,V(t)^\dag,\\
\label{completeW}
\paul{3}=V(t)^\dag\,\paul{3} V(t). 
\end{gather} Furthermore, we have derived a non-linear
equation of motion (\ref{collisionlesst}) for this entity in
Sec.~\ref{secDyn}.  It is not self-evident that the aforementioned constraints
on the structure of the density matrix are automatically preserved during the
time evolution.  Fortunately however, this is the case. This can be verified
easily by inserting Eq.~(\ref{DefGg}) into Eq.~(\ref{collisionlesst}). From
this, one finds that all instantaneous eigenvalues of the density matrix are
constants of motion \be P(t)=\text{const.}, \ee provided that the
instantaneous eigen basis evolves according to \be
i\hbar\frac{d}{dt}V(t)=\Sigma V \ee In turn, this is compatible with
Eq.~(\ref{completeW}) due to the symmetry of the self-energy in
Eq.~(\ref{sym2}).

\section{Results}
\label{secres}
\subsection{Rescaling the kinetic equations to a prolate, quasi
  one-dimensional configuration} General discussions on the properties of
many-body physics are usually plagued by a very abstract language. In order to
gain further insight into the complex non-linear physics, it is mostly
necessary to examine spatially homogeneous systems and assume stationary in
the time-domain.  This makes it possible to continue with analytical work and
to capture the essential bulk physics of macroscopic systems. However, trapped
quantum gases are different in many respects as the system size is of the same
order as the coherence length, thus boundary effects are of equal importance.
Moreover, it is difficult to discuss succinctly the meaning of the
thermodynamic limit in a quantum system with a mesoscopic particle number and
a discrete energy spectrum.

\begin{figure}[ht]
  \begin{center} 
    \includegraphics{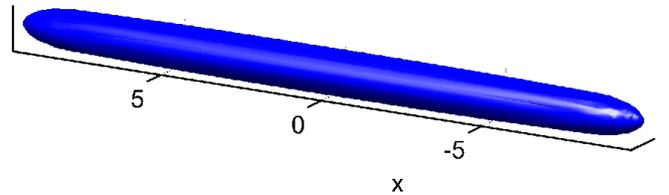}
  \caption{\label{cigar} Schematic representation of 
      the density distribution in a typical ``cigar-shaped'' trapped gas. The
      kinematic motion in the orthogonal directions $y,x$ is frozen out and
      presumably all the dynamics occurs along the weakly trapped $x$-axis.}
\end{center}
\end{figure}
In the following sections, we will therefore examine numerically the
properties of a quasi-one dimensional prolate system as shown in
Fig.~\ref{cigar}. While we have studied already a three-dimensional isotropic
configuration \cite{walser999}, prolate ``cigar-shaped'' traps are currently
at the focus of attention
\cite{goerlitz01,ertmer03,esslinger04,ott01}.  Due to the
reduction of the available phase-space volume, the role of quantum
fluctuations becomes more pronounced in low dimensional systems.

To be specific, we want to assume that the single particle trap Hamiltonian of
Eq.~(\ref{Hzero}) is of the form
\begin{gather}
  \mathcal{H}(\bx)=-\frac{\hbar^2}{2 m}\Delta+V_{\text{ho}}(\bx)-\mu,\\
  V_{\text{ho}}(\bx)=  \frac{1}{2} m {\omega}^2 x^2+
  \frac{1}{2} m {\omega_\perp}^2 (y^2+z^2). 
\end{gather}
A measure of the anisotropy of this trap geometry is the aspect ratio
$\beta=\omega_\perp/\omega$. In a very elongated, prolate configuration the
perpendicular oscillation frequency $\omega_\perp$ is much larger than the
longitudinal circular frequency $\omega$, thus $\beta \gg1$. In order to
reduce the kinetic equations also to a dimensionless form, we choose the
ground state extension of the longitudinal harmonic oscillator
$a_0=\sqrt{\hbar/m\,\omega}$ as the basic length scale, $t_0=2\pi/\omega$ as
the basic time-scale and $E_0=\hbar \,\omega$ as the natural energy unit.

The present formulation of the theory in terms of a generic set of basis
states $\{\ket{1}\}$ is able to handle the dimensional reduction from the
general three-dimensional kinetic theory to the quasi one-dimensional
situation quite easily, if we use an adapted trap basis in terms of
longitudinal and perpendicular harmonic oscillator states $\ket{1}=
\ket{i_1}\otimes\ket{j_1,k_1}_\perp$.  By assuming that only the ground state
components of the fields in the transverse directions are occupied, one can
limit the evolution to one dimension effectively
\begin{gather}
  \ket{\alpha}=\al{1}\ket{1} \otimes\ket{0,0}_\perp\\
  \fs{}=\fs{1,2}\ket{1} \bra{2} \otimes 
  \ket{0,0}_\perp \bra{0,0}_\perp,\\
  \ms{}=\ms{1,2}\ket{1} \ket{2} \otimes 
  \ket{0,0}_\perp \ket{0,0}_\perp.
\end{gather}
In order to evaluate the matrix elements of the binary interaction potential
further, we need the position representation of the normalized two-dimensional
harmonic oscillator ground-state
\begin{gather}
  \scal{y,z}{0,0}=\varphi_0(y,z;t)=e^{-i\beta \omega t}
  \sqrt{\frac{\beta}{\pi a_0^2 }}
  e^{-\beta \frac{y^2+z^2}{2 a_0^2}},\\
  \int dydz\,|\varphi_0|^2=1,\quad \int dydz \,|\varphi_0|^4=\frac{\beta}{2\pi
    a_0^2}.
\end{gather}
To compensate for the ground state energy of the two-dimensional harmonic
oscillator, we are working here with an explicitly time-dependent basis that
removes that energy shift.  We can now partially evaluate the matrix elements
of the two-body interaction in the contact potential approximation of
Eq.~(\ref{deltamatrixel}). By assuming only ground states in the transverse
direction, one obtains \bea
\label{deltamatrixelfac}
\phiunpr_0&=&\hbar\omega \, a_0 \frac{g}{2} \, \int dx\,
\scal{1}{x}\scal{2}{x}\scal{x}{3}\scal{x}{4}, \eea where we have introduced a
dimensionless coupling constant $g=2\beta\, a_s/a_0$.  This corresponds to an
effective one-dimensional interaction potential
$V_\text{bin}^{(1)}(x)=\hbar\omega a_0 g \,\delta(x)$, which would produce the
same matrix elements. In this limit of a very localized, point-like binary
interaction, one finds further that the self-energy operator of
Eqs.~(\ref{piGP}) and (\ref{sigmaHFB}) become local operators in space and
time , \ie, $\Pi(x,y,t)=\delta(x-y)\, \Pi(x,t)$ and $\Sigma(x,y,t)=\delta(x-y)
\,\Sigma(x,t)$.  Finally, this leads to the quasi one-dimensional
self-consistent Hartree-Fock-Bogoliubov (SCHFB) equations in the contact
potential approximation
\begin{align}
\label{piChi}
i\partial_t \chi&=\Pi(x,t)\chi(x,t),\\
\label{sigFluct}
i\partial_t G&=\Sigma(x_1,t)\,G(x_1,x_2,t)-G(x_1,x_2,t)  \Sigma(x_2,t)^\dag,
\end{align}
where local self-energies operators have been defined as follows
\begin{align}
\piGP{}(x,t)&=\left(
  \begin{array}{cc}
    \piGP{N}(x,t)& \piGP{A}(x,t)\\
    -\piGP{A}^\ast(x,t) & -\piGP{N}^\ast(x,t)
  \end{array}
\right),\\
 \piGP{N}(x,t)&=\frac{-\partial_{x}^2+x^2}{2}
+g\, |\alpha(x,t)|^2+2 g \,\fs{}(x,x,t)-\mu,\\
\piGP{A}(x,t)&=g \,\ms{}(x,x,t),
\end{align}
and 
\begin{align}
\Sigma(x,t)&=\left(
  \begin{array}{cc}
    \sigN(x,t)& \sigA(x,t)\\
    -\sigA^{\ast}(x,t) & -\sigN^\ast(x,t)
  \end{array}
\right),\\
\sigN(x,t)&=\frac{-\partial_{x}^2+x^2}{2}
+2 g \,|\alpha(x,t)|^2+2 g \,\fs{}(x,x,t)-\mu,\\
\sigA(x,t)&=g \,\alpha(x,t)^2+g\,\ms{}(x,x,t).
\end{align}

The numerical results that are presented in the following sections are based
on typical ${}^{87}$Rb condensate data, \ie, an atomic mass $m_{87}= 1.4431
10^{-25}$ kg, axial and radial trap frequencies $(\nu,\nu_\perp)=(3,800)$ Hz,
which result in an aspect ratio $\beta=267$, an axial harmonic oscillator
ground state size $a_0=6.2263\, \mu$m and a 3d scattering length
$a_s=5.8209$ nm. This leads to an effective coupling constant $g=2\beta
a_s/a_0=0.4986$.

\subsection{The Gross-Pitaevskii equation and the collective Bogoliubov
  excitations}
Before discussing the properties of the fully self-consistent equations, it is
prudent to start with the most basic version of it. By disregarding
$\fs{}$ and $\ms{}$ altogether, one obtains the Gross-Pitaevskii
equation
\begin{gather}
\label{equGP}
  i\partial_t\chi=\Pi^\text{GP}(x,t)\,\chi(x,t),\\
  \piGP{N}^{\text{GP}}(x,t)= \frac{-\partial_{x}^2+x^2}{2}+g\,
  |\alpha(x,t)|^2-\mu, \; \piGP{A}^{\text{GP}}=0.
\end{gather}
From a stationary solution $\chi_0=(\al{0},\ald{0})^\top$, which is normalized
to the total particle number $N=\int dx |\al{0}(x)|^2$, one obtains the
chemical potential $\mu(N)$. In the interaction dominated mean-field regime,
one may disregard the kinetic energy contribution altogether and one finds in
the Thomas-Fermi approximation $\mu_{\text{TF}}=(3g N/2)^{2/3}/2$, the width
of the condensate $x_\text{TF}=\sqrt{2 \mu_\text{TF}}$ and a healing length
$\xi=1/x_\text{TF}$.

\subsubsection{Linear response analysis} 
From a weak perturbation around the stationary solution $\alpha(x,t)=e^{-it
  \,\delta\mu}[\al{0}(x)+\delta\alpha(x,t)]$ or equivalently \be
\label{bogmod}
\chi(x,t)=e^{-it\, \paul{3} \delta\mu }
\left[ \chi_0(x)+\delta\chi(x,t) \right], \ee one
obtains the collective Bogoliubov excitations modes and linear response
frequencies of Eq.~(\ref{equGP}).
\begin{align}
\label{sigBog}
i\partial_t\delta\chi=&
 \Sigma^{\text{B}}(x) \delta\chi(x,t)-\delta\mu \,P,\\
\sigN^{\text{B}}(x)=&\frac{-\partial_{x}^2+x^2}{2}+2 g \,|\alpha_0(x)|^2-\mu,\\
\sigA^{\text{B}}(x)=&g \,\alpha_0(x)^2,
\end{align}
The energy spectrum for the linear response matrix $\Sigma^\text{B}$ must be
gap-less ($E_0=0$) as the GP-Eq.~(\ref{equGP}) is $U(1)$ invariant under a
global phase change of $\al{0}$. In other words, there exists a degenerate
manifold of ground state solutions and it requires no energy to transform one
of them into another. The infinitesimal generator of this phase rotation is
the zero-mode $P=(\al{0},-\ald{0})^\top$ and it can be verified easily that
$\Sigma^\text{B}P=0$.

The initial perturbation $\delta\chi$ can also induce a small change in the
particle number $\delta N=P^\dag\paul{3}\delta\chi$.  In turn, this
leads to a small modifications of the chemical potential $\mu(N+\delta
N)=\mu(N)+\delta\mu$ at which the perturbed system is evolving globally. By
taking this into account in the ansatz for the perturbation analysis in
Eq.~(\ref{bogmod}), one is able to cope with the secular terms that arise
otherwise in a gap-less linear response analysis
\begin{align}
  \delta\chi(t)&=e^{-it\,
    \Sigma^\text{B}} \delta\chi(x,0) +it\, \delta\mu \,P\nonumber\\
  &=iQ \,\delta N -iP\,\delta\Phi+ W^\prime e^{-i
    t\,E^\prime}\paul{3}^{(n-1)}\delta\chi^\prime,
\end{align}
where we have used the completeness relation of Eq.~(\ref{completeness}) and
defined Bogoliubov amplitudes as $\delta\Phi=Q^\dag\paul{3}\delta\chi$, as
well as $\delta\chi^\prime={W^\prime}^\dag\paul{3}\delta\chi$.  The residual
energy shift of the chemical potential is proportional to the inverse of the
inertial mass of the collective ground state mode $\delta\mu=\delta N/M$.
Within the TF approximation, one finds for the mass parameter
$M_\text{TF}=(12gN)^{1/3}/g$.

\begin{figure}[t]
  \begin{center} 
    \includegraphics[height=\columnwidth,angle=-90]{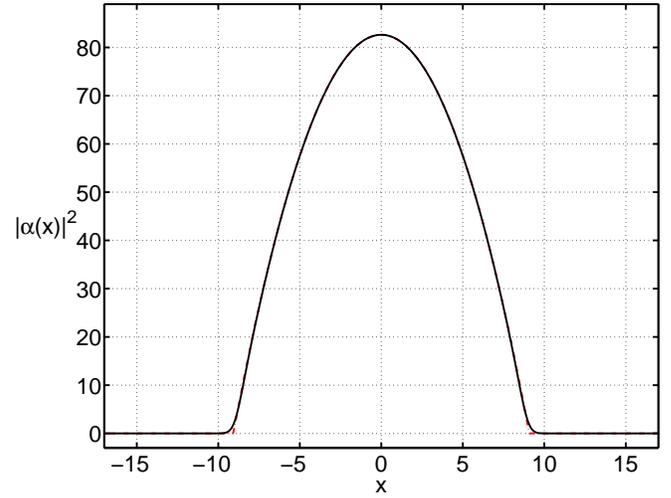}
   \caption{\label{GPTF} The condensate density $|\alpha(x)|^2$ as a function 
     of the position $x$ in units of the harmonic oscillator length for a
     particle number of $N=10^3$.  Only in the proximity of the Thomas-Fermi
     radius $x_\text{TF}=9.07$ the exact numerical solution (solid line) is
     distinguishable from the Thomas-Fermi approximation (dashed doted line).
     The chemical potential energy in h.o. units $\mu = 41.205 $ is also well
     approximated by $\mu_{\text{TF}}=41.198$.}
\end{center}
\end{figure}
In considering the similarities of the collective Bogoliubov excitations
Eq.~(\ref{sigBog}) and the quasi-particle modes of the generalized self-energy
matrix of Eq.~(\ref{sigFluct}), it is of utmost importance not to confuse
their different physical meaning.  Quasi-particle modes address the questions
of excitations of the quantum fluctuations above a static mean-field without
considering the back-action.  Consequently, they do not have to be gap-less
nor satisfy a Kohn theorem \cite{dobson94}, but they are just a convenient
basis to describe the quantum vacuum or the thermal excitations thereof. Thus,
they should not be considered as better or worse approximation of each other.
Only the linear response analysis of the coupled system of
Eqs.~(\ref{piChi}, \ref{sigFluct}) will be comparable to the collective
Bogoliubov excitations of Eq.~(\ref{sigBog}). It is straight forward to verify
that the coupled system of Eqs.~(\ref{piChi},\ref{sigFluct}) are $U(1)$
invariant, thus gap-less $(E_0=0)$. Moreover, the collective center of mass
oscillation also decouples from the internal excitations and it evolves exactly
with the harmonic oscillator frequency of $E_1=1$.
\begin{figure}[t]
  \begin{center} 
    \includegraphics[height=\columnwidth,angle=-90]{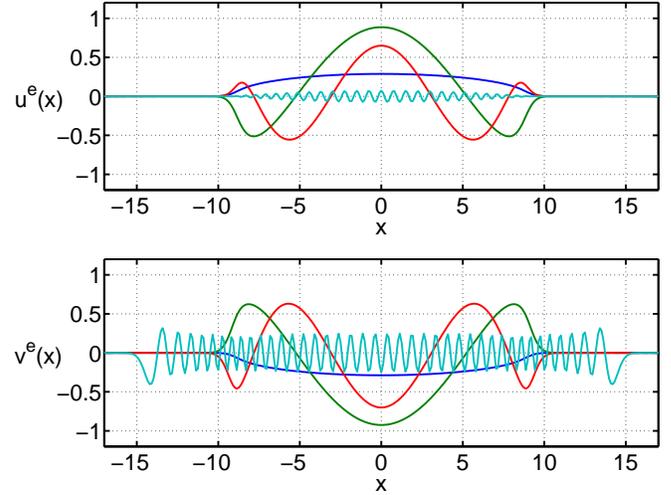}
   \caption{\label{Boge} The normalized, even spatial components of the 
     negative energy Bogoliubov excitation mode $W(E_n<0)=[u_e,v_e^\ast]^\top$
     as a function of the position $x$ in units of the harmonic oscillator
     length.  In particular, we show the three low energy modes
     $E_{n=(0,2,4)}=(0,-1.732, -3.167)$ and an arbitrarily chosen higher
     energetic mode $E_{80}=-63.845$ for comparison.}
\end{center}
\end{figure}
\begin{figure}[ht]
  \begin{center} 
    \includegraphics[height=\columnwidth,angle=-90]{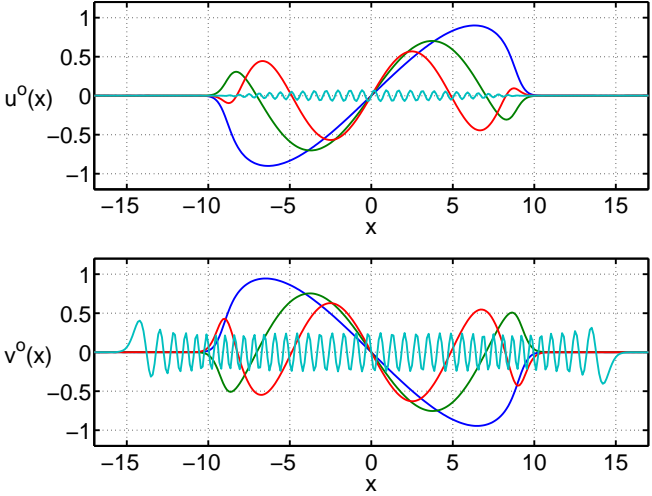}
   \caption{\label{Bogo} The normalized, odd  spatial components of the 
     negative energy Bogoliubov excitation mode $W(E_n<0)=[u_o,v_o^\ast]^\top$
     as a function of the position $x$.  Depicted are the three low energy
     modes $E_{n=(1,3,5)}=(-1, -2.451, -3.882)$ and an arbitrarily chosen
     higher energetic mode $E_{81}=-64.723$ for comparison.}
\end{center}
\end{figure}
\subsubsection{Static mean-field and collective Bogoliubov excitations 
  for $10^3$ particles}
In Figs.~\ref{Boge} and \ref{Bogo}, we depict a few selected
collective excitation modes of the GP-equation corresponding to the static
solution with $N=10^3$ particles shown in Fig.~\ref{GPTF}. The reflexion
symmetry of the harmonic trap is carried over to the mean-field state.
Consequently, one can also classify the collective Bogoliubov modes according
to even and odd parity modes. Conventionally, one introduces also hole $u(E<0)$
and particle $v(E<0)$ amplitudes as components of
$W(x,E_n<0)=[u(x,E),v^\ast(x,E)]^\top$.  The particle character of $v(E<0)$
becomes visible in the higher energy excitation modes $E_{80,81}$ where only
the particle-like amplitudes reach far outside the spatial extend of the
condensate and oscillate between the classical turning points of fictitious
particles with energy $|E|$. In contrast, hole-like excitations always remain
localized on the site of the condensate wave-function and their amplitudes
decrease with decreasing excitation energy.

\subsubsection{Chemical potential and collective Bogoliubov excitation
      energies for a full range of particle numbers}
\begin{figure}[t]
  \begin{center} 
    \includegraphics[height=\columnwidth,angle=-90]{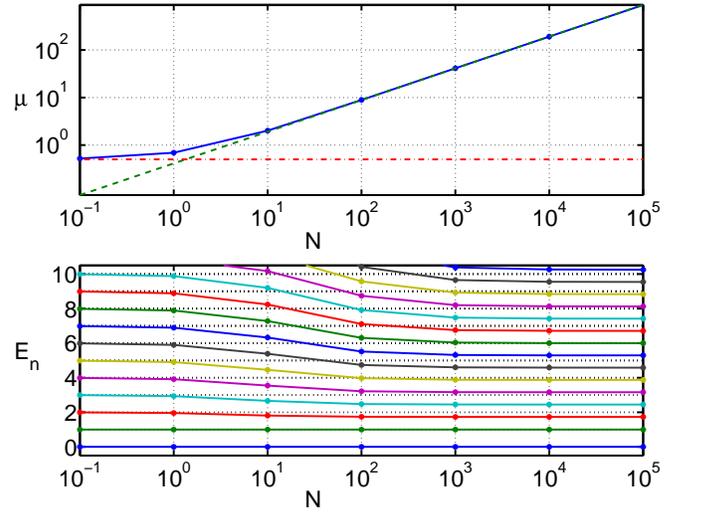}
    \caption{\label{BogN} The upper part shows  
      the chemical potential $\mu(N)$ of the GP equation as a function of
      particle number (solid line). For comparison, we present also the Thomas
      Fermi approximation $\mu_\text{TF}(N)$ (dashed line) and the harmonic
      oscillator ground state energy of $1/2$ (dashes doted line). In the
      lower part, the collective Bogoliubov energies $E$ are shown versus
      particle number. All units are given in h.o. energies.}
\end{center}
\end{figure}
The general character of the solutions depends parametrically on the particle
number. In Fig.~\ref{BogN}, we have plotted the chemical potential $\mu(N)$ as
well as the collective Bogoliubov energies $E(N)$ as a function of the
particle number $N$. The double logarithmic representation of the chemical
potential shows clearly a linear slope, hence reproduces the power law
dependence that is found within the Thomas-Fermi approximation
$\mu_{\text{TF}}(N)=(3g N/2)^{2/3}/2$. The collective excitation frequencies
are identical to the harmonic oscillator spectrum for small particle numbers.
However with increasing particle number, the energy density of states visibly
increases as the level spacing is reduced. By focusing on the two lowest
energy excitations, \ie, the Goldstone mode $E_0=0$ and Kohn mode $E_1=1$, one
finds that those fundamental symmetry generating modes are unaffected by the
interactions.

\subsection{Ground state solutions of the self-consistent 
Hartree-Fock-Bogoliubov equations}
In this section, we will discuss the self-consistent solution of the
stationary Eqs.~(\ref{piChi}) and (\ref{sigFluct}), \ie,
\begin{align}
0&=\Pi\chi,\quad
0=\Sigma\,G-G  \Sigma^\dag.
\end{align}
In particular, we are interested in the lowest energy configuration of the
system. Thus, no quasi-particle modes shall be occupied and according to
Eq.~(\ref{genericgmat}), we construct the quantum vacuum only from the
negative energy states $W_{-}=W_{E<0}$ with dim[$W_{-}$]=$2n\times n$ \be G=
W_{-}^{\phantom{\dag}} W_{-}^\dag. \ee With an iterative procedure, one can
solve for the mean-field amplitude $\al{0}$ assuming static $\fs{}$ and
$\ms{}$. In a second step, one has to find the quasi-particle modes and
construct a density matrix from it. This procedure has to be continued until
convergence is reached. We have computed the self-consistent solutions for a
range of particle numbers $N=(10^0, 10^1, 10^2, 10^3, 10^4,10^5)$. For a
particle number $N<1$, this extended mean-field theory is neither physically
meaningful, nor did it lead to converging solutions any longer.

\subsubsection{Static mean-field and quasi-particle modes for $10^3$ 
  particles} 
\begin{figure}[t]
  \begin{center} 
    \includegraphics[height=\columnwidth,angle=-90]{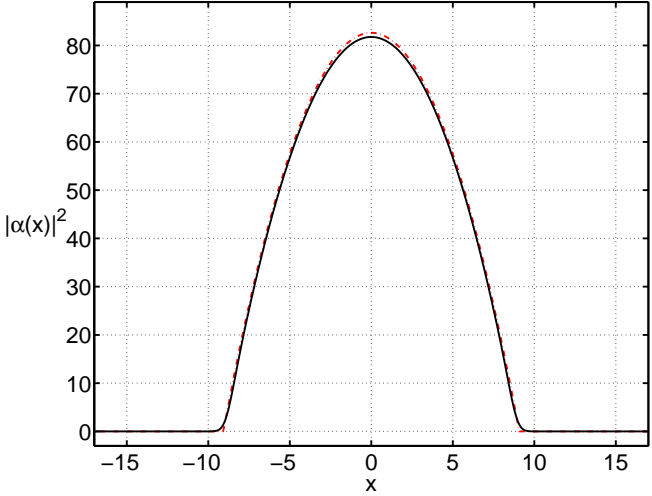}
   \caption{\label{scGPTF} The self-consistent mean-field density 
     $|\alpha(x)|^2$ (solid line) and Thomas-Fermi approximation (dashed doted
     line) as a function of the position $x$ in units of the h.o. length.}
\end{center}
\end{figure}
For a particle number $N=10^3$, we show in Fig.~\ref{scGPTF} the mean-field
density as a function of position. Due to a repartitioning of particles
between condensate and the non-condensed fraction, there are now fewer
particles in the mean-field component. Thus, the difference to the
Thomas-Fermi approximation is more visible than in Fig.~\ref{GPTF}.  For a
total particle number of $N=10^3=N^{(c)}+\tilde{N}$, we find a
$(N^{(c)},\tilde{N}) =(984.83,15.17)$ and chemical potential of $\mu = 40.61$
in h.o.  energy units.
\begin{figure}[t]
  \begin{center} 
    \includegraphics[height=\columnwidth,angle=-90]{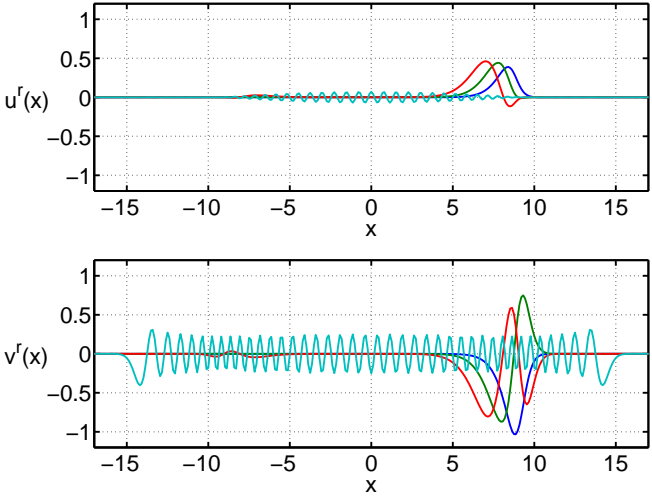}
   \caption{\label{scBogr} The normalized, spatial components of the 
     negative energy Bogoliubov excitation mode
     $W(E_n<0)=[u_{r},v_{r}^\ast]^\top$ as a function of the position $x$ in
     units of the harmonic oscillator length.  In particular, we show the
     three low energy modes $E_{n=(0,2,4)}=( -3.356, -6.421, -8.636)$ and an
     arbitrarily chosen higher energetic mode $E_{80}= -64.819$ for
     comparison. The low energy modes are localized on the right side of the
     condensate and almost degenerate to the modes localized on the other side
     depicted in Fig.~\ref{scBogl}.}
\end{center}
\end{figure}

\begin{figure}[t]
  \begin{center} 
    \includegraphics[height=\columnwidth,angle=-90]{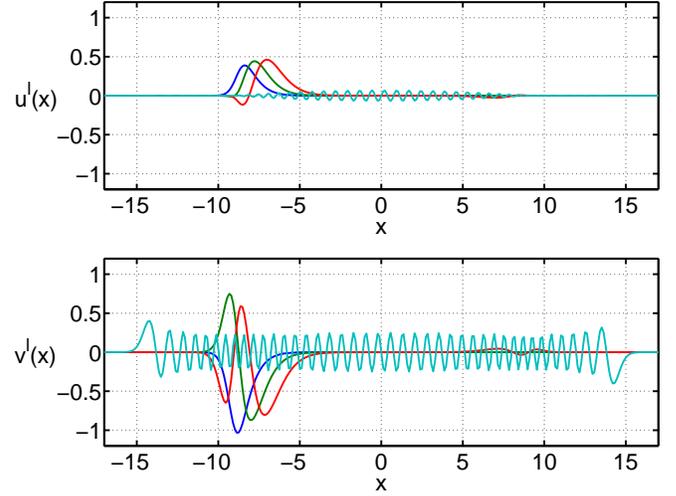}
   \caption{\label{scBogl} The normalized, spatial components of the 
     negative energy Bogoliubov excitation mode
     $W(E_n<0)=[u_{l},v_{l}^\ast]^\top$ as a function of the position $x$.
     Depicted are the three low energy modes $E_{n=(1,3,5)}=(-3.3556, -6.421,
     -8.636)$ and an arbitrarily chosen higher energetic mode $E_{81}=-65.689$
     for comparison. The low energy modes are localized on the left side of
     the condensate.}
\end{center}
\end{figure}

The quasi-particle modes are depicted in Figs.~\ref{scBogr} and \ref{scBogl}.
While the collective Bogoliubov modes of Figs.~\ref{Boge} and \ref{Bogo} can
be characterized with a definite even or odd parity, this is seemingly not the
case here.  The low energy quasi-particle modes are localized on left and
right sides of the condensate and have more of the character of the single
particle Hartree-Fock excitations discussed in Eqs.~(\ref{Hsq}) and
(\ref{sigHF}), where the potential energy $V_{\text{ho}}+2\,\Uc-\mu$ has a
double minimum at $\pm x_\text{TF}$. However, one has to take into account
also that the low lying modes are energetically degenerate and that the higher
energy excitations do exhibit a definite parity.  Thus, one could construct
quasi-particle modes with definite parity by symmetrizing or anti-symmetrizing
them. We have deliberately chosen not do so in order to permit the occurrence
of a reflexion symmetry breaking. For example, spatial symmetry breaking is
observed in deformed nuclei \cite{blaizot} or the mixing of two component Bose
gases \cite{Esry1998b}.  Nevertheless, we did not find such a symmetry
breaking behavior here.
\subsubsection{Chemical potential and quasi-particle energies for a full range
  of particle numbers} 
\begin{figure}[t]
  \begin{center} 
    \includegraphics[height=\columnwidth,angle=-90]{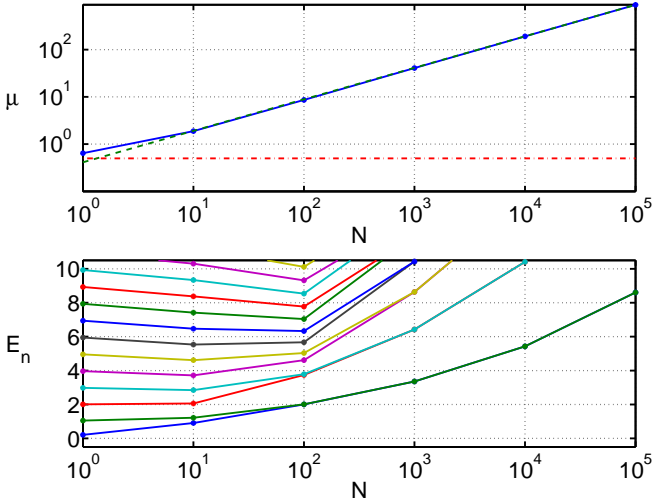}
    \caption{\label{scBogN} Upper part: 
      the chemical potential $\mu(N)$ (solid line) of the SCHFB equations, the
      TF approximation $\mu_\text{TF}$ (dashed line) and the h.o. ground state
      energy $1/2$ (dashed dotted line) versus particle number $N$.  Lower
      part: quasi-particle energy spectrum $E(N)$ versus particle number $N$.}
\end{center}
\end{figure}

In Fig.~\ref{scBogN}, we present the chemical potential
$\mu(N)$ as well as the quasi-particle energies $E$ as function of the total
particle number $N$.  One can see very clearly the energy gap $E_0\neq0$ and
the absence of the Kohn mode $E_1\neq 1$, which should occurs only for the
collective excitations of system as whole. Moreover, one can observe how the
low lying quasiparticle modes begin to coalesce with gradually increasing
particle number.
\begin{figure}[t]
  \begin{center} 
    \includegraphics{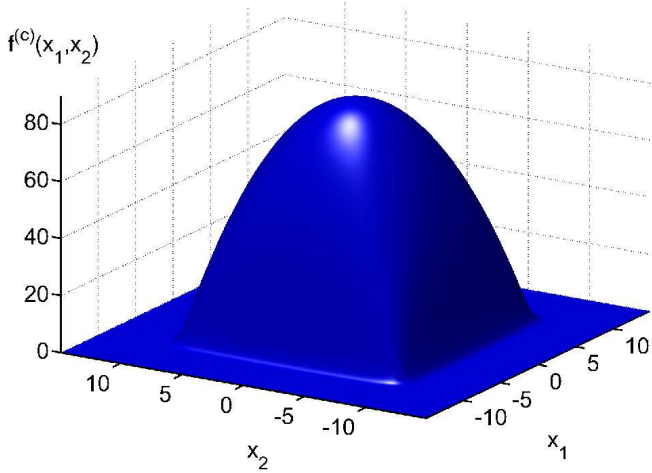}
\caption{\label{fc} Coherent contribution of the single particle 
  density matrix $\fc{}(x_1,x_2)$ as a function of the spatial coordinates for
  a particle number of $N=10^3$. The coherence extends in the diagonal as well
  as in off-diagonal directions up to the TF radius. For a real valued
  mean-field $\alpha(x)$, the coherent contribution of the pairing field
  $\mc{}(x_1,x_2)$ is also represented by this figure.}
\end{center}
\end{figure}
\subsection{The single particle density matrix and the pairing field}
The mean-field amplitude $\alpha(x)$, the total single particle density
$\ft{}(x_1,x_2)=\fc{}(x_1,x_2)+\fs{}(x_1,x_2)$, as well as the pairing field
$\ms{}(x_1,x_2)$ have been the central concepts of the present analysis. Thus,
we will present in the following pictures instances of their spatial
representation for a particle number of $N=10^3$.  This gives a good
qualitative impression of the universal features of the ground state.  Results
obtained for different particle numbers are similar in appearance and we will
discuss the quantitative differences next.

\subsubsection{Spatial representation for $10^3$ particles}
In Fig.~\ref{fc}, we show the coherent contribution to single particle density
matrix $\fc{}(x_1,x_2)=\alpha^\ast(x_2)\,\alpha(x_1)$. As it is constructed
from the order parameter $\alpha(x)$, it has the full ODLRO
\cite{onsager56,yang62}, which extends over the complete system size.  For the
one-dimensional trap that we consider in here, the Hamiltonian operator is
real valued, thus the ground state solution of the mean-field $\alpha(x)$ is a
purely real quantity, too.  Consequently,
$\mc{}(x_1,x_2)=\alpha(x_2)\,\alpha(x_1)$ is identical to $\fc{}(x_1,x_2)$ as
depicted in Fig.~\ref{fc}.

\begin{figure}[t]
  \begin{center} 
     \includegraphics{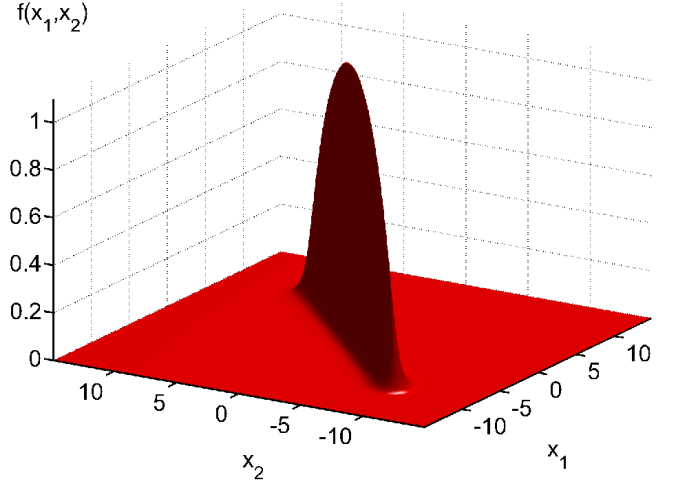}
    \caption{\label{fsq} Non-condensate density matrix $\fs{}(x_1,x_2)$ as a
      function of the spatial coordinates for a particle number of $N=10^3$.}
\end{center}
\end{figure}

In contrast to the full ODLRO which is found in the mean-field component, one
finds that the non-condensate density $\fs{}(x_1,x_2)$ is predominantly
localized along the coordinate diagonal and exhibits spatial variation only
due to the external confinement with a trapping potential. In the off-diagonal
direction this order parameter has only a very short range which is determined
by the binary interaction. This is depicted in Fig.~\ref{fsq}.

\begin{figure}[t]
  \begin{center} 
 \includegraphics{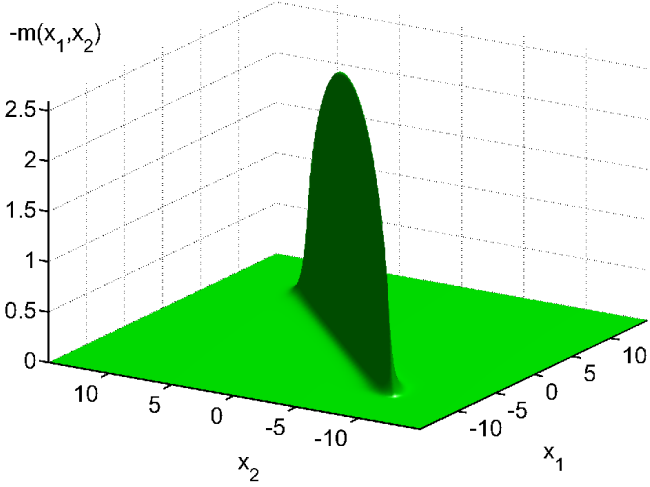}
    \caption{\label{msq} The negative pairing field $-\ms{}(x_1,x_2)$ 
      as a function of the spatial coordinates for a particle number of
      $N=10^3$.}
\end{center}
\end{figure}

Interestingly, one finds also for the pairing field $\ms{}(x_1,x_2)$ a very
similar spatial behavior.  However, while a single particle interpretation is
sufficient to understand the behavior of the normal density matrix, it is
necessary to use two-particle physics in the pairing field of Fig.~\ref{msq}.
The strong negative correlation along the diagonal show that there is a
reduced likelihood of finding two particles at the same location. Again, this
likelihood is modulated by the density of particles in trap.  However, the
degree of anti-correlation drops off quickly with increasing distance between
the positions at which particles are extracted.

\subsubsection{Diagonal and off-diagonal elements of $\fs{}$ and $\ms{}$
  for a full range of particle numbers}

\begin{figure}[t]
  \begin{center} 
    \includegraphics[height=\columnwidth,angle=-90]{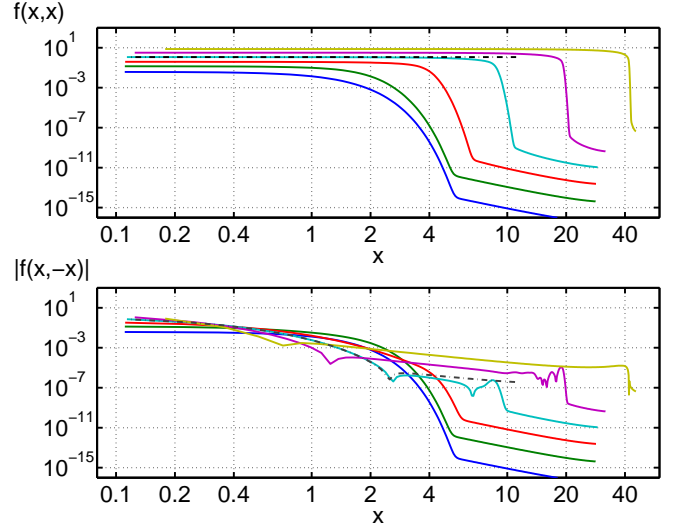}
    \caption{\label{diagfsq} Diagonal $\fs{}(x,x)$ and 
      off-diagonal $|\fs{}(x,-x)|$ elements of the normal density matrix
      versus distance $x$ for a full range of particle numbers
      $N=(10^0,\ldots,10^5)$. The individual results can be identified easily
      by the spatial extension that grows proportional with the particle
      number. The dashed-dotted line shows the result for a homogeneous
      gas corresponding to a trapped gas with $N=10^3$ particles.}
\end{center}
\end{figure}

\begin{figure}[t]
  \begin{center} 
    \includegraphics[height=\columnwidth,angle=-90]{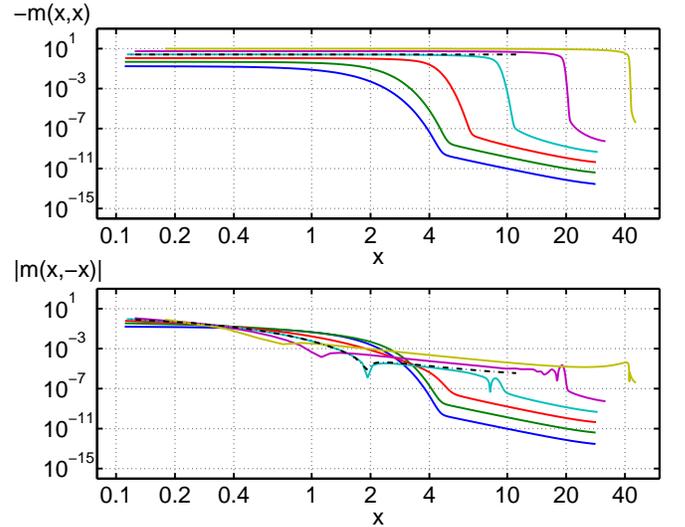}
    \caption{\label{diagmsq} 
      Diagonal $-\ms{}(x,x)$ and off-diagonal $|\ms{}(x,-x)|$ elements of
      the pairing field versus distance $x$ as in Fig.~\ref{diagfsq}.}
\end{center}
\end{figure}

In this section, we have compiled the quantitative results for the spatial
variation of the diagonal $\fs{}(x,x)$, $\ms{}(x,x)$ and the off-diagonal
$\fs{}(x,-x)$, $\ms{}(x,-x)$ elements of the normal density matrix and the
pairing field, respectively, for a full range of particle numbers
$N=(10^0,\ldots,10^5)$. The double logarithmic representation used in
Figs.~\ref{diagfsq} and \ref{diagmsq} reveals clearly that there is a
separation of the bulk physics in the center of the trap and the physics
dominated by the boundary at the rim of the condensate.  We have verified this
separation of scales by turning of the trap potential. In this case, one
recovers the homogeneous limit.  In order to make a quantitative comparison
with the trapped system, we have chosen a homogeneous mean-field density
$n=N/2L=82.62$ that matches the mean-field density at the center of the
trapped gas for a particle number of $N=10^3$. This leads to very similar
chemical potentials for the homogeneous system of $\mu_\text{hom}=40.47$ and
for the trapped gas of $\mu_\text{trap}=40.61$, respectively.  The dashed
dotted lines in Figs.~\ref{diagfsq} and \ref{diagmsq}, do represent the
homogeneous results and compare very well with the trapped gas for $N=10^3$.
This proves that the local density approximation (LDA) yields a good
approximation for the transverse correlation length. A detailed comparison of
the critical exponents and their dependence on temperature is currently under
investigation.

\subsection{Effective coupling constants}
\begin{figure}[t]
  \begin{center} 
    \includegraphics[height=\columnwidth,angle=-90]{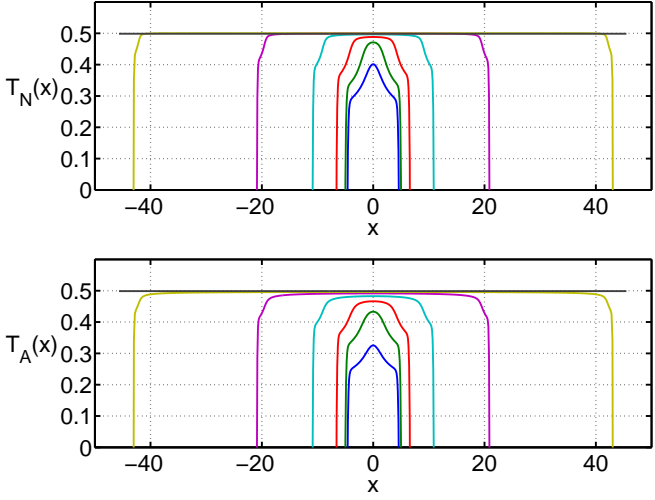}
    \caption{\label{Tren} The effective coupling constants 
      $T_{\mathcal{N}}(2\mu,x)$ and $T_{\mathcal{A}}(2\mu,x)$ as a function of
      position. Each subplot contains six individual lines corresponding to
      the full range of particle numbers $N=(10^0, \ldots, 10^5)$ and a flat
      line, which the represents the bare coupling constant $g=0.4986$. The
      individual results can be identified easily by their spatial extension,
      which increases proportional to the TF radius or particle number,
      respectively.}
  \end{center}
\end{figure}

In Sec.~\ref{secTmat}, we have discussed formally the renormalization of the
binary interaction to an effective T-matrix, which takes place in any 
self-consistent calculation. In particular, we upgraded the anomalous
potential of Eq.~(\ref{TA}) and related it to an anomalous many-body T-matrix.
As we have calculated the SCHFB equations in here, we can now revisit this
question and find out what the effective renormalized coupling constant is
\begin{align}
T_{\mathcal{A}}(2\mu,x)\, \mc{}(x,x)&=g\,[\mc{}(x,x)+\ms{}(x,x)],\\
T_{\mathcal{N}}(2\mu,x)\, \fc{}(x,x)&=2g\,[\fc{}(x,x)+\fs{}(x,x)].
\end{align}
From the numerical results of Fig.~\ref{Tren}, we can draw three conclusions:
a) the effective coupling is position (or momentum) dependent and decreases
rapidly outside of the range of the condensate; b) the effective coupling
constant is in general less than the bare interaction constant, which may be
understood in terms of a second order perturbation theory; c) the effective
coupling constant at the center of the trap becomes gradually less for smaller
particle numbers.

\subsection{First and second order correlations functions}
\subsubsection{Spatial representation for $10^3$ particles}
Quantum fluctuations around the classical mean-field amplitude are the central
topic of this article. In the previous sections, we have examined the single
particle density $\ft{x}=\fc{}+\fs{}$, which is directly an observable
quantity and the pair correlation function $\ms{}(x_1,x_2)$, which is not.
The questions of how to quantify and to measure quantum correlations has
always been a central theme for any quantized field theory, whether in
condensed matter physics \cite{onsager56,yang62,Naraschewski996} or in quantum
optics \cite{hanburybrowntwiss56,glauber63}.

In essence, first order coherence is measured by the correlation function 
\begin{align}
  g^{(1)}(x_1,x_2)&=\frac{ \langle \aopd{x_2} \aop{x_1} \rangle}{\sqrt{
      n(x_1)\,n(x_2)}}\\
  &=\frac{\fc{}(x_1,x_2)+\fs{}(x_1,x_2)}{ \sqrt{ n(x_1)\,n(x_2)}},\nonumber
\end{align}
where $n(x)=f^{(c)}(x,x)+\tilde{f}(x,x)$ denotes the total density.
Primarily, it is sensitive only to spatial phase correlations or, in other
words, off-diagonal order. This can bee seen easily by disregarding the
quantum depletion $\fs{}$ for the moment. In the case of a mean-field
$\alpha(x)=|\alpha(x)| e^{-i\Phi(x)}$, it is only proportional to the phase
gradients $\Phi(x)=\Phi(0)+\delta\Phi(x)$. Moreover, if $\alpha(x)$ is a
static ground state without a phase gradient (irrotational), one finds the
definition of full coherence, \ie, $g^{(1)}(x_1,x_2)=1$.

\begin{figure}[t]
  \begin{center} 
  \includegraphics{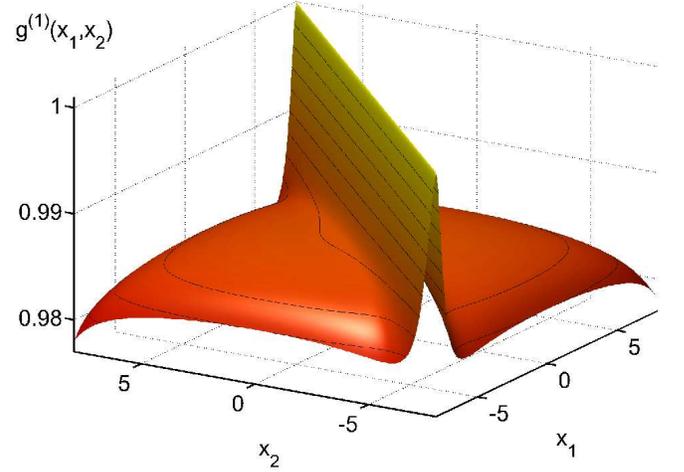}
    \caption{\label{gone} First order correlation function $g^{(1)}(x_1,x_2)$
      versus spatial position for $10^3$ particles.}
\end{center}
\end{figure}

In Fig.~\ref{gone}, we have evaluated the first order correlation function for
a trapped gas with $N=10^3$ particles. By definition, it must be exactly 1
along the diagonal. As expected, there is only a very weak influence of the
quantum fluctuations noticeable, since $\fc{}(x_1,x_2)\gg\fs{}(x_1,x_2)$,
which can be seen explicitly in Fig.~\ref{scGPTF} and Fig.~\ref{fsq}.
Consequently, one finds that first order coherence is not a very sensitive
probe for the quantum aspects of a field. For example, sending a classical
optical field through an semi-opaque or noisy medium immediately leads to a
reduction coherence, which is as such a purely classical phenomenon.

\begin{figure}[t]
  \begin{center} 
    \includegraphics{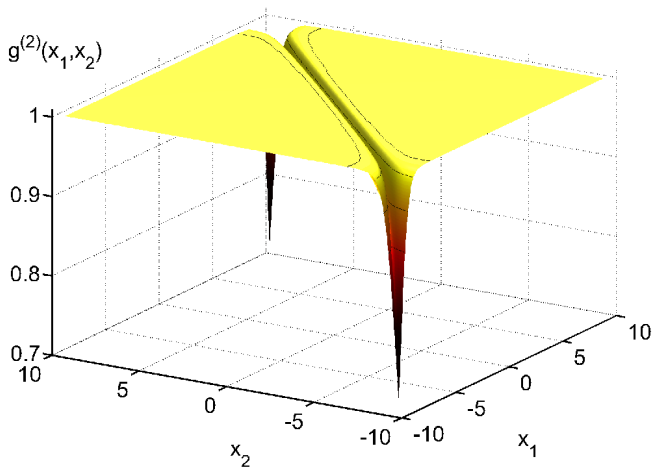}
    \caption{\label{gtwo} Second order correlation function $g^{(2)}(x_1,x_2)$
      versus spatial position for $10^3$ particles.}
\end{center}
\end{figure}

Actually, a sensitive probe for the quantum nature of a field is the second
order correlation function $g^{(2)}$. If $g^{(1)}$ basically responds to
uncertainty in the phase quadrature, then $g^{(2)}$ is affected by density
fluctuations $\aopd{x}\aop{x}=n(x)+\delta\hat{n}(x)$. Intuitively speaking,
this is the conjugate variable to the phase gradient. More succinctly
speaking, this can be examined by studying a Heisenberg uncertainty product
(see App.~\ref{CauchySchwartz}) for all the field quadratures as in
Eq.~(\ref{quasiparticles}).  Explicitly, it is defined as the normal ordered
density-density correlation function:
 \begin{align}
   &g^{(2)}(x_1,x_2)=\frac{
     \langle  \aopd{{x_1}} \aopd{{x_2}} \aop{{x_2}} \aop{{x_1}} \rangle}{
     n({x_1}) n(x_{2})}=
   1+\frac{1}{n({x_1}) n(x_{2})} \Bigr{\{}\nonumber\\
   &+2\, \Re{\left[\fc{}(x_1,x_2)^\ast 
   \fs{}(x_2,x_1) +
   \mc{}(x_1,x_2)^\ast
   \ms{}(x_2,x_1)\right]}\nonumber\\
   &+\fs{}(x_1,x_2) \fs{}(x_2,x_1)+
     \ms{}(x_1,x_2)^\ast \ms{}(x_2,x_1)\Bigl{\}},
\end{align}
which is shown in Fig.~\ref{gtwo} for $10^3$ particles.  While $g^{(2)}$ it is
mostly equal to 1, it can be seen clearly that along the diagonal
$g^{(2)}(x,x)<1$. This is a unique signature a non-classical state of the
quantum field with a sub-poisonian statistic of number fluctuations or, in
other words, a number squeezed state \cite{Orzel01}. The suppression of number
fluctuations has its origin the binary interaction potential of
Eq.~(\ref{Hoppos}).  By reducing the local density fluctuations, one can
remove more energy than what is regained by the kinetic energy due to the
increased phase uncertainty. At its extreme this is also the physical
mechanism that leads to the Mott phase transition
\cite{jaksch98,bloch02,Greiner2002b,rwalser1002} where on-site interactions
compete with nearest neighbor tunneling.

\subsubsection{Diagonal and off-diagonal elements of the first and second order
  correlation function for a full range of particle numbers}
 
\begin{figure}[t]
  \begin{center} 
    \includegraphics[height=\columnwidth,angle=-90]{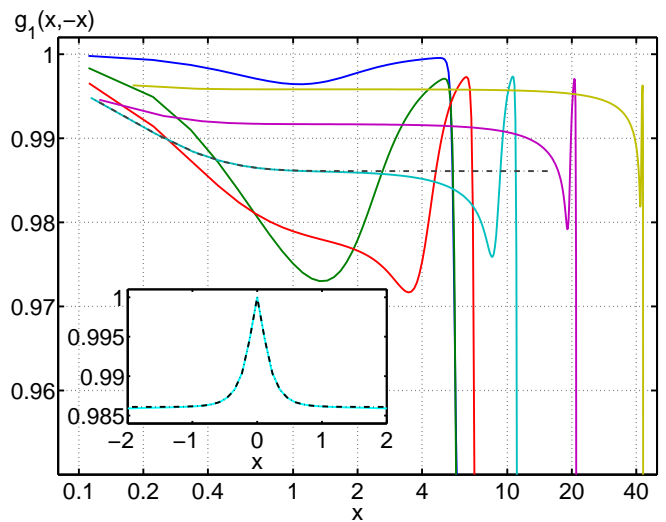}
    \caption{\label{goneall} Off-diagonal elements of the
      first order correlation function $g^{(1)}(x,-x)$ versus spatial
      position.  Each subplot contains six individual lines corresponding to
      the full range of particle numbers $N=(10^0,\ldots,10^5)$ and a dashed
      dotted line, which represents the homogeneous gas result. The individual
      results can be identified easily by their spatial extension, which
      increases proportional to the particle number.}
\end{center}
\end{figure}

\begin{figure}[t]
  \begin{center} 
    \includegraphics[height=\columnwidth,angle=-90]{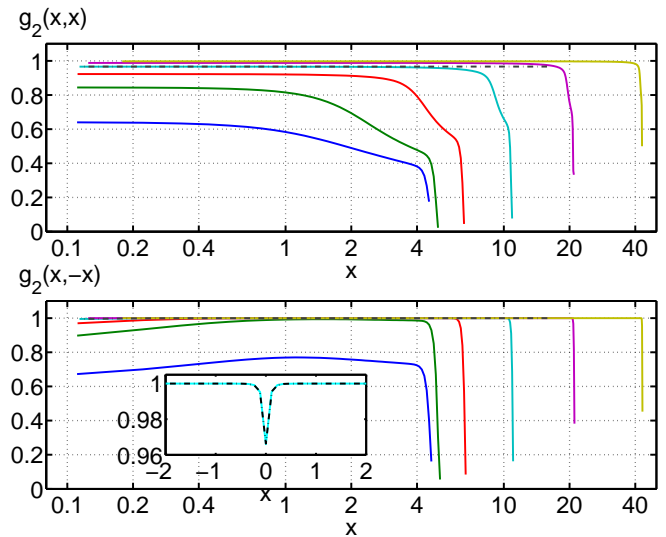}
    \caption{\label{gtwoall} Diagonal $g^{(2)}(x,x)$ and off-diagonal  
      $g^{(2)}(x,-x)$ elements of the second order correlation function versus
      spatial position. Parameters and legend as in Fig.~(\ref{goneall}).}
\end{center}
\end{figure}

Finally, we summarize the results for the first and second order correlation
function in Figs.~\ref{goneall} and \ref{gtwoall} for the full range of
particle numbers $N=(10^0,\ldots,10^5)$. As before, we have also computed the
result for a homogeneous gas in the local density approximation, which
corresponds to a trapped gas of $10^3$ particles.  The insets in the pictures
magnify both of those curves in the central region where they do agree very
well.  However, at the rim of the condensate the trapped gas does exhibit
features that are absent in the LDA.

It is physically most important to see in Figs.~\ref{gtwoall} that the
suppression density fluctuations becomes stronger for smaller particle
numbers. At its limit, this anti-correlation of tow bosonic particles leads to
an effective fermionization and is the hall mark of the Tonks-Girardeau regime
\cite{girardeau60,bloch04,shlyapnikov04}.  If this argument is applied to the
situation of an inhomogeneous trapped gas, then this means that the number
squeezing is larger at the rim than in the center of the cloud.

\section{Conclusions and Outlook}
\label{concout}
In this article, we have given a basic overview of the premises and concepts
of a number-symmetry broken non-equilibrium kinetic theory of a trapped
bosonic gas. By extending the successful mean-field concept of the
Gross-Pitaevskii equation with the effects of non-local, two particle quantum
correlations, one obtains a renormalized binary interaction T-matrix and
allows for the dynamic establishment of non-classical many-particle quantum
correlations. At very low temperatures, we can disregard the equilibrating
effects of elastic two-particle collisions, in contrast to previous work
\cite{walser599,walser999,wachter501}. In this limit, we have proven that the
important physical constants of motion, like particle number, energy or the
entropy are conserved. Obviously, the inclusion of collision processes is
desirable from a fundamental point of view, as this will break the
micro-reversibility and lead to a thermal relaxation towards the most probable
distribution as dictated by thermodynamics. In preliminary study
\cite{bhongale}, we have already studied the
effects of collisions and the consequences of various approximations in the
absence of a mean-field in a model system, but further work is necessary.
 
In the second section of the paper, we have focused on the specific properties
of the zero-temperature ground state correlations of a trapped, quasi-one
dimensional bosonic gas.  This maximizes the role played by quantum
fluctuations, due to the dimensional reduction of the available phase-space
volume. With a fully self-consistent numerical calculation, we have evaluated,
the mean-field amplitudes, the quantum depletion, the pairing field as well as
the first and second order ground state quantum correlations for a full range
of particle numbers $N=(10^0,\ldots,10^5)$. Most interestingly, we do get a
strong suppression of the density fluctuations or, in other words, an enhanced
number squeezing with decreasing particle density. This generic feature is in
general agreement with the predictions that are found with exactly solvable
one-dimensional models such as the Tonks-Girardeau gas or the Bose-Hubbard gas
on an optical lattice. A detailed analytical comparison with those models
\cite{shlyapnikov04,andersen02,alkhawaja02} and the experimental results
\cite{ertmer03,esslinger04} is currently work in progress. In conclusion, we
find that the general non-equilibrium kinetic theory also reproduces the
ground state correlations of a quasi one-dimensional bosonic gas well.


\section*{Acknowledgments} 
R.~W. acknowledges gratefully many stimulating discussions with J. Cooper and
F. Sols as well as travel support from the ESF through the BEC2000 program.
\appendix
\section{Cauchy-Schwartz inequality}
\label{CauchySchwartz}
For  a positive semi-definite density operator $\sigma$
and an arbitrary  operator $\hat{L}$ it 
follows that the expectation value
\bea
\label{genposexp}
\av{\hat{L}\, \hat{L}^\dag}&=& \text{Tr}\{\sigma\,\hat{L}\,
\hat{L}^\dag\}\ge 0 \eea is never negative.  Consequently, the
co-variance matrix $G$ of Eq.~(\ref{chi}) must be positive
semi-definite $u^\dag G u \ge 0$, as well.  This can be easily seen,
by considering a linear combination of two arbitrary operators
$\hat{A}$ and $\hat{B}$, \ie, $L=\alpha \hat{A}+\beta
\hat{B}$.  By minimizing the positive expression
Eq.~(\ref{genposexp}), one obtains the Cauchy-Schwartz inequality as
\bea \av{\hat{A}\hat{A}^\dag}\av{\hat{B}\hat{B}^\dag}\ge
\av{\hat{B}\hat{A}^\dag}\av{\hat{A}\hat{B}^\dag}.  \eea In particular,
for the special choice of $\hat{A}=\delta\aop{1}$ and
$\hat{B}=\delta\aopd{2}$, this implies that the magnitude of the
anomalous fluctuations is limited by \bea (1+\fs{11})\fs{22}\ge
|\ms{12}|^2.  \eea

\section{Pauli matrices}
\label{appa}
In this article, we use the following standard representation for the Pauli
matrices: \bea \paul{1}= \left({\bf
    \begin{array}{cc}
      0& \mathds{1}\\
      \mathds{1} & 0
    \end{array}}
\right), \paul{2}= i\,\left({\bf
    \begin{array}{cc}
      0& -\mathds{1}\\
      \mathds{1} & 0
    \end{array}}
\right), \paul{3}= \left({\bf
    \begin{array}{cc}
      \mathds{1}& 0\\
      0 & -\mathds{1}
    \end{array}}
\right),\nonumber\\ \eea They satisfy the ordinary commutation relation of an
angular momentum operator $ [\paul{1},\paul{2}]=i \paul{3} $ and all cyclic
permutations thereof.  If $n$ is the dimension of the vector space, then
$\paul{k}$ acts in a $2n$-dimensional symplectic vector space \cite{blaizot}.

\section{Canonical transformations}
\label{canonical}   
A canonical transformation is an inhomogeneous linear combination of
creation and destruction operators that preserves the commutation
relation \cite{blaizot}.  In particular, if $\aop{}$ and
$\aopd{}$ denotes a pair of hermitian conjugated bosonic operators,
such that
\bea [ \aop{1},\aopd{2} ]&=&\delta_{1,2}, \eea 
then any affine linear transformation defines a new set of 
operators $b$ and 
$\bar{b}$ by 
\bea
\label{affine}
\left(
  \begin{array}{c}
     b\\
    \bar{b}
  \end{array}
\right)
&=& T\,\left(
  \begin{array}{c}
    \aop{}\\
    \aopd{}
  \end{array}
\right)+d.  \eea In an $n$-dimensional vector space, $T$ represents a
$2\,n\times2\,n$ dimensional matrix and $d$ is a $2\,n$ dimensional
vector.  Such a transformation is canonical if the new pair of operators
also satisfies the commutation relation:  \bea
\label{commutb}
[ b_{1},\bar{b}_{2}]=\delta_{1,2}.  \eea More specifically, the
transformation is unitary canonical if the new operators are hermitian
conjugate pairs, \ie, $\bar{b}=b^\dag$.  By inserting
Eq.~(\ref{affine}) into Eq.~(\ref{commutb}), one finds that the
transformation matrices are a representation of the symplectic group
$Sp(2\,n)$: \bea T\,\paul{3}\, T^\dag=\paul{3}. \eea In addition, it
can be shown that $T^{\ast}=\paul{1}\,T\,\paul{1}$ and
$T^{-1}=\paul{3}\,T^{\dag}\paul{3}$.  

\section{Quantum limit for the ground state correlations}
\label{appb} 
The state of the interacting many-body system is described within the set of
approximations by a mean-field amplitude $\al{}$ and a generalized $G$ matrix.
At finite temperature or out of equilibrium, there is a finite occupation
number of particles of positive energy excited states. However, in the lowest
energy configuration all allocatable real particles occupy the mean-field
amplitude and the generalized density matrix only holds the vacuum.
Then, the following idem-potency relation holds for the density
matrix
\begin{align}
  G\, \paul{3}\, G+G=0.
\end{align}
This follows straight from Eq.~(\ref{genericgmat}) when $P_+=0$.
Consequently, this gives a restriction for the components of the density
matrix $\fs{},\ms{}$. In particular, one finds that the vacuum depletion is
completely determined by the pairing field
\begin{gather}
 \fs{}=\frac{1}{2}(
 \sqrt{\mathds{1}+4\,\ms{} {\ms{}}^\dag}-\mathds{1}),\\
  \fs{}\ms{}=\ms{}\fs{}^\ast.
\end{gather}
It is important to note that this relation also holds out-of-equilibrium as
long as thermalizing collisions can be disregarded.

\section{A generalized Wick's theorem}
\label{wick}
Gau{\ss}ian fluctuations around a well defined mean value are the key
assertion to apply Wick's theorem \cite{peletminskii}.  This is a set of rules to
efficiently evaluate quantum averages for multiple operator products as \bea
\label{multop}
\aval{ \kpsiop{1} \kpsiop{2} \ldots \kpsiop{l}}
 \eea In this average,
for example, the operator $\kpsiop{1}$ represents either an operator
$\aop{1}$ or $\aopd{1}$.

First, the displacement rule shifts any operator $\kpsiop{1}$ by its
c-number expectation value $\psi_{1}$ which is either $\al{1}$ or
$\ald{1}$, and replaces the quantum average by an average that has
zero mean values: \bea \lefteqn{ \aval{ \kpsiop{1} \kpsiop{2}\ldots
    \kpsiop{l}}=}\\
&=&\avnull{(\kpsiop{1}+\psi_{1}) (\kpsiop{2}+\psi_{2})\ldots
  (\kpsiop{n}+\psi_{l})}.\nonumber \eea Second, after expanding the
multiple products, one can discard all averages that involve an odd
numbers of operators: \bea \avnull{ \kpsiop{1} \kpsiop{2}\ldots
  \kpsiop{2s+1}}&=&0.  \eea And third, for the remaining averages, one
can use the Gau{\ss}ian factorization rule: \bea \lefteqn{ \avnull{
    \kpsiop{1} \kpsiop{2}\ldots
    \kpsiop{2s}}=}\\
&=&\avnull{ \kpsiop{1} \kpsiop{2}} \avnull{ \kpsiop{3}\ldots
  \kpsiop{2s}}+\nonumber\\
&+&\avnull{ \kpsiop{1} \kpsiop{3}} \avnull{ \kpsiop{2} \kpsiop{4}
  \ldots
  \kpsiop{2s}}+\nonumber\\
&&\quad\quad\vdots\nonumber\\
&+&\avnull{ \kpsiop{1} \kpsiop{2s}} \avnull{ \kpsiop{2} \ldots
  \kpsiop{2s-1}}.\nonumber \eea By proceeding recursively, one has
finally evaluated the complete multiple operator average
Eq.~(\ref{multop}).
      
\bibliographystyle{osa} \bibliography{MyPublications,bec}

\begin{thebibliography}{10}

\bibitem{stringarireview}
F. Dalfovo, S. Giorgini, L. Pitaevskii, and S. Stringari, ``Theory of trapped
  Bose-condensed gases,'' Rev. Mod. Phys. {\bf 71,} 463 (1999).

\bibitem{leggett401}
A. Leggett, ``Bose-Einstein condensation in the alkali gases: Some fundamental
  concepts,'' Rev. Mod. Phys. {\bf 73,} 307 (2001).

\bibitem{southwell}
K. Southwell, ``Ultra cold matter,'' Nature {\bf 416,} 205 (2002).

\bibitem{pethick02}
C. Pethick and H. Smith, {\em Bose-Einstein Condenstion in Dilute Gases}
  (Cambridge University Press, 2002).

\bibitem{cornellsci}
{M. H. Anderson and J. R. Ensher and M. R. Matthews and C.~E.~Wieman and
  E.~A.~Cornell}, ``Observation of Bose-Einstein Condesation in a Dilute Atomic
  Vapour,'' Science {\bf 269,} 198 (1995).

\bibitem{ketterle1195}
K.~B. Davis, M.-O. Mewes, M.~R. Andrews, N.~J. van Druten, D.~S. Durfee, D.~M.
  Kurn, and W. Ketterle, ``Bose-Einstein condensation in a gas of sodium
  atoms,'' Phys. Rev. Lett. {\bf 75,} 3969 (1995).

\bibitem{hulet895}
C.~C. Bradley, C.~A. Sackett, J.~J. Tollett, and R.~G. Hulet, ``Evidence of
  Bose-Einstein Condensation in an Atomic Gas with Attractive Interactions,''
  Phys. Rev. Lett. {\bf 75,} 1687 (1995).

\bibitem{holland999}
J. Williams and M. Holland, ``Preparing topological states of a Bose-Einstein
  condensate,'' Nature {\bf 401,} 568 (1999).

\bibitem{matthews99}
M. Matthews, B. Anderson, P. Haljan, D. Hall, C. Wieman, and E. Cornell,
  ``Vortices in a Bose-Einstein Condensate,'' Phys. Rev. Lett. {\bf 83,} 2498
  (1999).

\bibitem{ketterle0401}
J. Abo-Shaeer, C. Raman, J. Vogels, and W. Ketterle, ``Observation of Vortex
  Lattices in Bose-Einstein Condensates,'' Science {\bf 292,} 476 (2001).

\bibitem{jaksch98}
D. Jaksch, C. Bruder, J. Cirac, C. Gardiner, and P. Zoller, ``Cold bosonic
  atoms in optical lattices,'' Phys. Rev. Lett. {\bf 81,} 3108 (1998).

\bibitem{bloch02}
M. Greiner, O. Mandel, T. Esslinger, T. H{\"a}nsch, and I. Bloch, ``Quantum
  phase transitions from a superfluid to a Mott insulator in a gas of ultracold
  atoms,'' Nature {\bf 415,} 39 (2002).

\bibitem{Greiner2002b}
M. Greiner, O. Mandel, T.~W. H{\"a}nsch, and I. Bloch, ``Collapse and revival
  of the matter wave field of a {B}ose-{E}instein condensate,'' Nature {\bf
  419,} 51 (2002).

\bibitem{regal03}
C. Regal, C. Ticknor, J. Bohn, and D. Jin, ``Creation of ultracold molecules
  from a Fermi gas of atoms,'' Nature {\bf 424,} 47 (2003).

\bibitem{zwierlein04}
M. Zwierlein, C. Stan, C. Schunck, S. Raupach, A. Kerman, and W. Ketterle,
  ``Condensation of Pairs of Fermionic Atoms near a Feshbach Resonance,'' Phys.
  Rev. Lett. {\bf 92,} 120403 (2004).

\bibitem{regal04}
C. Regal, M. Greiner, and D. Jin, ``Observation of Resonance Condensation of
  Fermionic Atom Pairs,'' Phys. Rev. Lett. {\bf 92,} 040403 (2004).

\bibitem{chin04}
C. Chin, M. Bartenstein, A. Altmeyer, S. Riedl, S. Jochim, J.~H. Denschlag, and
  R. Grimm, ``Observation of the Pairing Gap in a Strongly Interacting Fermi
  Gas,'' Science {\bf 305,} 1128 (2004).

\bibitem{bloch04}
B. Paredes, A. Widera, V. Murg, O. Mandel, S. F\"olling, I. Cirac, G.~V.
  Shlyapnikov, T.~W. H\"ansch, and I. Bloch, ``Tonks-Girardeau gas of ultracold
  atoms in an optical lattice,'' Nature {\bf 429,} 277 (2004).

\bibitem{das02}
K. Das, M. Girardeau, and E. Wright, ``Crossover from One to Three Dimensions
  for a Gas of Hard-Core Bosons,'' Phys. Rev. Lett. {\bf 89,} 110402 (2002),
  and references therein.

\bibitem{onsager56}
O. Onsager and L. Penrose, ``Bose-Einstein Condensation and Liquid Helium,''
  Phys. Rev {\bf 104,} 576 (1956).

\bibitem{yang62}
C. Yang, ``Concept of Off-Diagonal Long-Range Order and the Quantum Phases of
  Liquid He and of superconductors,'' Rev. Mod. Phys. {\bf 34,} 694 (1962).

\bibitem{beliaev58b}
S. Beliaev, ``Energy-Spectrum of a Non-Ideal Bose Gas,'' JETP {\bf 34,} 299
  (1958).

\bibitem{Hohenberg1965}
P. Hohenberg and P. Martin, ``Microscopic Theory of Superfluid Helium,'' Ann.
  Phys. {\bf 34,} 291 (1965).

\bibitem{zaremba899}
E. Zaremba, T. Nikuni, and A. Griffin, ``Dynamics of trapped bose gases at
  finite temperatures,'' J. Low. Temp. Phys. {\bf 116,} 277 (1999).

\bibitem{rusch599}
M. Rusch and K. Burnett, ``Mean-field theory of excitations of trapped Bose
  condesates at finite temperatures,'' Phys. Rev. A {\bf 59,} 3851 (1999).

\bibitem{Stoof1999a}
H.~T.~C. Stoof, ``Coherent Versus Incoherent Dynamics During {B}ose-{E}instein
  Condensation in Atomic Gases,'' J. Low Temp. Phys. {\bf 114,} 11 (1999).

\bibitem{walser599}
R. Walser, J. Williams, J. Cooper, and M. Holland, ``Quantum kinetic theory for
  a condensed bosonic gas,'' Phys.~Rev.~A {\bf 59,} 3878 (1999).

\bibitem{wachter501}
J. Wachter, R. Walser, J. Cooper, and M. Holland, ``Equivalent kinetic theories
  of Bose-Einstein condensation,'' Phys.~Rev.~A {\bf 64,} 053612 (2001).

\bibitem{giorgini00}
S. Giorgini, ``Collisionless dynamics of dilute Bose gases: Role of quantum and
  thermal fluctuations,'' Phys. Rev. A {\bf 61,} 63615 (2000).

\bibitem{jackson02}
B. Jackson and E. Zaremba, ``Quadrupole collective modes in trapped
  finite-temperature Bose-Einstein condensates,'' Phys. Rev. Lett. {\bf 88,}
  180402 (2002).

\bibitem{bhongale}
S. Bhongale, R. Walser, and M. Holland, ``Memory effects and conservation laws
  in the quantum kinetic evolution of a dilute Bose,'' Phys.~Rev.~A {\bf 66,}
  043618 (2002).

\bibitem{Naraschewski996}
M. Naraschewski and R.~J. Glauber, ``Spatial coherence and density correlations
  of trapped Bose gases,'' Phys. Rev. A {\bf 59,} 4595 (1999).

\bibitem{esslinger02}
I. Bloch, T. H{\"a}nsch, and T. Esslinger, ``Measurement of the spatial
  coherence of a trapped Bose gas at the phase transition,'' Nature {\bf 403,}
  166 (2000).

\bibitem{Orzel01}
C. Orzel, A. Tuchman, M. Fenselau, and M.~K. M.~Yasuda, ``{Squeezed States in a
  Bose-Einstein Condensate},'' Science {\bf 291,} 2386 (2001).

\bibitem{mermin66}
N.~D. Mermin and H. Wagner, ``Absence of Ferromagnetism or Antiferromagnetism
  in One- or Two-Dimensional Isotropic Heisenberg Models,'' Phys. Rev. Lett.
  {\bf 17,} 1133 (1966).

\bibitem{hohenberg67}
P.~C. Hohenberg, ``Existence of Long-Range Order in One and Two Dimensions,''
  Phys. Rev. {\bf 158,} 383 (1967).

\bibitem{shlyapnikov04}
D. Petrov, D. Gangardt, and G. Shlyapnikov, ``Low-dimensional trapped gases,''
  cond-mat/0409230  (2004), and references therein.

\bibitem{andersen02}
J. Andersen, U.~A. Khawaja, and H. Stoof, ``Phase Fluctuations in Atomic Bose
  Gases,'' Phys. Rev. Lett. {\bf 88,} 70407 (2002).

\bibitem{alkhawaja02}
U.~A. Khawaja, J. Andersen, N. Proukakis, and H. Stoof, ``Low dimensional Bose
  gases,'' Phys. Rev. A {\bf 66,} 13615 (2002).

\bibitem{ertmer03}
D. Hellweg, L. Cacciapuoti, M. Kottke, T. Schulte, K. Sengstock, W. Ertmer, and
  J.~J. Arlt, ``Measurement of the Spatial Correlation Function of Phase
  Fluctuating Bose-Einstein Condensates,'' Phys. Rev. Lett. {\bf 91,} 010406
  (2003).

\bibitem{Olshanii98}
M. Olshanii, ``Atomic Scattering in the Presence of an External Confinement and
  a Gas of Impenetrable Bosons,'' Phys. Rev. Lett. {\bf 81,} 938 (1998).

\bibitem{menotti02}
C. Menotti and S. Stringari, ``Collective oscillations of a one-dimensional
  trapped Bose-Einstein gas,'' Phys. Rev. A {\bf 66,} 043610 (2002).

\bibitem{esslinger04}
T. St{\"o}ferle, H. Moritz, C. Schori, M. K{\"o}hl, and T. Esslinger,
  ``Transition from a Strongly Interacting 1D Superfluid to a Mott Insulator,''
  Phys. Rev. Lett. {\bf 92,} 130403 (2004).

\bibitem{verhaar694}
A.~J. Moerdijk and B.~J. Verhaar, ``Prospects for Bose-Einstein condensation in
  atomic 7Li and 23Na,'' Phys. Rev. Lett. {\bf 73,} 518 (1994).

\bibitem{wieman495}
N.~R. Newbury, C.~J. Myatt, and C.~E. Wieman, ``s-wave elastic collisions
  between cold ground-state 87Rb atoms,'' Phys. Rev. A {\bf 51,} 2680 (1995).

\bibitem{morgan02}
S. Morgan, M. Lee, and K. Burnett, ``Off-shell T matrices in one, two, and
  three dimensions,'' Phys. Rev. A {\bf 65,} 22706 (2002).

\bibitem{huang}
K. Huang, {\em Statistical Mechanics} (John Wiley \& Sons, Inc., New York,
  1965).

\bibitem{lepage97}
P. Lepage, ``How to renormalize the Schr{\"o}dinger equation,'' xxx.lanl.gov
  nucl-th/9706029 (1997).

\bibitem{exactsolutions}
 in {\em Exactly Solvable Problems in Condensed Matter and Relativistic Field
  Theory}, Vol.~242 of {\em Lecture Notes in Physics}, H. Araki, ed.,
  (Springer, 1985).

\bibitem{girardeau60}
M. Girardeau,  J. Math. Phys. {\bf 1,} 516 (1960).

\bibitem{richardson68}
R. Richardson, ``Exactly solvable Many-Boson Model,'' J. Math. Phys. {\bf 9,}
  1327 (1968).

\bibitem{dukelsky01}
J. Dukelsky and P. Schuck, ``Condensate fragmentation in a new exactly solvable
  model for confined bosons,'' Phys. Rev. Lett. {\bf 86,} 4207 (2001).

\bibitem{schoenhammer96}
K. Sch{\"o}nhammer and V. Meden, ``Fermion-boson transmutation and comparison
  of statistical ensembles in one dimension,'' Am. J. Phys. {\bf 64,} 1168
  (1996).

\bibitem{chapman}
S. Chapman and T.~G. Cowling, {\em The Mathematical Theory of Non-Uniform
  Gases} (Cambridge University Press, Cambridge, 1970).

\bibitem{schuck}
P. Ring and P. Schuck, {\em The Nuclear Many-Body Problem} (Springer Verlag,
  N.Y., 1980).

\bibitem{peletminskii}
A.~I. Akhiezer and S.~V. Peletminskii, {\em Methods of Statistical Physics}
  (Pergamon Press Ltd., Oxford, England, 1981).

\bibitem{abrikosov65}
A. Abrikosov, L. Gor'kov, and I. Dzyaloshinskii, {\em Quantum field theoretical
  methods in statistical physics} (Pergamon Press, Oxford, England, 1965).

\bibitem{blaizot}
J.~P. Blaizot and G. Ripka, {\em Quantum Theory of Finite Systems} (The MIT
  Press, Cambridge, Massachusetts, 1986).

\bibitem{schrieffer}
J.~R. Schrieffer, {\em Theory of Superconductivity} (Perseus Books, Cambridge,
  Massachusetts, 1999).

\bibitem{schleich01}
W.~P. Schleich, {\em Quantum optics in Phase Space} (Wiley-VCH, Berlin,
  Germany, 2001).

\bibitem{griffin496}
A. Griffin, ``Conserving and gapless approximations for an inhomogeneous Bose
  gas at finite temperatures,'' Phys. Rev. B {\bf 53,} 9341 (1996).

\bibitem{girardeau59}
M. Girardeau and R. Arnowitt, ``Theory of Many-Boson Systems: Pair Theory,''
  Phys. Rev. {\bf 113,} 755 (1959).

\bibitem{Castin1998a}
Y. Castin and R. Dum, ``Low-temperature {B}ose-{E}instein condensates in
  time-dependent traps: {B}eyond the {\it{U}}(1) symmetry-breaking approach,''
  Phys. Rev. A {\bf 57,} 3008 (1998).

\bibitem{zollerv}
C.~W. Gardiner and P. Zoller, ``Quantum kinetic theory V: Quantum kinetic
  master equation for mutual interaction of condensate and noncondensate,''
  Phys. Rev. A {\bf 61,} 033601 (2000), and references therein.

\bibitem{kadanoff62}
L. Kadanoff and G. Baym, {\em Quantum Statistical Mechanics}, {\em Frontiers in
  Physics} (W. A Benjamin, Inc., New York, 1962).

\bibitem{kadanoff65}
J. Kane and L. Kadanoff, ``Green's Functions and Superfluid Hydrodynamics,'' J.
  of Math. Phys. {\bf 6,} 1902 (1965).

\bibitem{golub96}
G. Golub and C.~V. Loan, {\em Matrix Computations} (Johns Hopkins Unversity
  Press, Baltimore, 1996).

\bibitem{morgan04}
S. Morgan, ``Response of Bose-Einstein condensates to external perturbations at
  finite temperature,'' Phys. Rev. A {\bf 69,} 023609 (2004).

\bibitem{Lewenstein1996a}
M. Lewenstein and L. You, ``Quantum Phase Diffusion of a {B}ose-{E}instein
  Condensate,'' Phys. Rev. Lett. {\bf 77,} 3489 (1996).

\bibitem{You1998}
L. You, W. Houston, M. Lewenstein, and M. Marinescu, ``Low energy excitations
  of a trapped Bose condensates,'' Act. Phys. Pol. A {\bf 93,} 211 (1998).

\bibitem{walser999}
R. Walser, J. Cooper, and M. Holland, ``Reversible and irreversible evolution
  of a condensed bosonic gas,'' Phys.~Rev.~A {\bf 63,} 013607 (2001).

\bibitem{goerlitz01}
A. G{\"o}rlitz, J. Vogels, A. Leanhardt, C. Raman, T.~G. ad~J.~Abo-Shaeer, A.
  Chikkatur, S. Gupta, S. Inouye, T. Rosenband, and W. Ketterle, ``Realization
  of Bose-Einstein Condensates in Lower Dimensions,'' Phys. Rev. Lett. {\bf
  87,} 130402 (2001).

\bibitem{ott01}
H. Ott, J. Fortagh, G. Schlotterbeck, A. Grossmann, and C. Zimmermann,
  ``Bose-Einstein Condensation in a Surface Microtrap,'' Phys. Rev. Lett. {\bf
  87,} 230401 (2001).

\bibitem{dobson94}
J. Dobson, ``Harmonic-Potential Theorem: Implications for Approximate Many-Body
  Theories,'' Phys. Rev. Lett. {\bf 73,} 2244 (1994).

\bibitem{Esry1998b}
B.~D. Esry and C.~H. Greene, ``Superfluids mixing it up,'' Nature {\bf 392,}
  434 (1998).

\bibitem{hanburybrowntwiss56}
R.~H. Brown and R. Twiss, ``Correlation between photons in two coherent beams
  of light,'' Nature {\bf 177,} 27 (1956).

\bibitem{glauber63}
R. Glauber, ``The Quantum Theory of Optical Coherence,'' Phys. Rev. {\bf 130,}
  2529 (1963).

\bibitem{rwalser1002}
R. Walser, ``Quantenfelder $\grave{\text{a}}$ la carte,'' Physik Journal {\bf
  11,} 19 (2002).

\end{thebibliography}

\end{document}